\documentclass[11pt]{article}
\usepackage[utf8]{inputenc}
\usepackage{graphicx}
\graphicspath{ {images/} }
\usepackage{amsmath,amssymb,amsthm,textcomp}
\usepackage{titling}
\usepackage{bbm}
\usepackage{caption}
\usepackage{tikz,pgfplots}
\usepackage[margin=1in]{geometry}
\usepackage{multirow,booktabs}
\usepackage [english]{babel}
\usepackage [autostyle, english = american]{csquotes}
\MakeOuterQuote{"}

\newtheorem{definition}{Definition}

\usepackage{setspace}
\usepackage{tikz}
\usepackage{subfig}
\usepackage[mathscr]{euscript}
\usetikzlibrary{matrix}
\usepackage{tcolorbox}
\definecolor{mycolor}{rgb}{0.122, 0.435, 0.698}

\newtcbox{\mybox}{on line,
  colframe=mycolor,colback=mycolor!10!white,
  boxrule=0.5pt,arc=4pt,boxsep=0pt,left=6pt,right=6pt,top=6pt,bottom=6pt}
\usepackage[hyphens,spaces]{url} 
\providecommand{\keywords}[1]
{
  \small	
  \textbf{\textit{Keywords---}} #1
}
\usepackage[colorlinks,citecolor=blue,urlcolor=blue,bookmarks=false,hypertexnames=true]{hyperref} 
\usepackage{apacite}
\title{\textbf{\LARGE{Errors in Learning from Others' Choices}}}
\usepackage{geometry}
\author{
\large{Mohsen Foroughifar}\thanks{Rotman School of Management, University of Toronto, ON, Canada, mohsen.foroughifar@rotman.utoronto.ca. This research has been approved by the University of Toronto Research Ethics Board under Human Participant Ethics Protocol \#36847. Financial Support from the BEAR center (Behavioural Economics in Action at Rotman) is gratefully acknowledged. I am grateful to Ryan Webb for supervising this project. I would like to thank Yoram Halevy, Tanjim Hossain, Yesim Orhun, David Soberman, Sandro Ambuehl, Guillaume Fréchette, Ben Greiner, Deborah Small, and participants of the Marketing Brown Bag at Rotman, BEAR Summer Research Retreat, and  Marketing Science Conference for helpful feedback. Johannes Hoelzemann and Pavandeep Singh kindly helped with the logistics of the experiment.}
}

\begin{document}
\maketitle

\begin{abstract}
Observation of other people's choices can provide useful information in many circumstances. However, individuals may not utilize this information efficiently, i.e., they may make decision-making errors in social interactions. In this paper, I use a simple and transparent experimental setting to identify these errors. In a within-subject design, I first show that subjects exhibit a higher level of irrationality in the presence than in the absence of social interaction, even when they receive informationally equivalent signals across the two conditions. A series of treatments aimed at identifying mechanisms suggests that a decision maker is often uncertain about the behavior of other people so that she has difficulty in inferring the information contained in others' choices. Building upon these reduced-from results, I then introduce a general decision-making process to highlight three sources of error in decision-making under social interactions. This model is non-parametrically estimated and sheds light on what variation in the data identifies which error.
%


\end{abstract}

\newpage

\section{Introduction}

Understanding the interactions of economic agents is a central concern in the economics of information \shortcite{Manski00}. Learning from others' choices is important in many circumstances: patients draw negative quality inferences from the refusal of a kidney in the kidney market \shortcite{Zhang10}; observation of a neighbor's choice in a presidential election reveals information about his/her preferences \shortcite{OrhunUrminsky13}; editors of scientific journals accept/reject papers on the basis of referee decisions; traders in financial markets infer information about asset fundamental values from the order flow, among others. In these contexts, behavioral errors may have severe consequences and ultimately can lead to a socially inefficient outcome, e.g., poor kidney utilization despite the continual shortage in kidney supply. The goal of this paper is to identify the sources of such errors.

%
%
%
%
The fundamental question in social interactions is that how people "glean" information from others' choices. Econometric analysis of choice data often reduces the empirical inference to revelation of preferences by assuming that individuals are Bayesian and have rational expectations. These assumptions are typically necessary for identification of beliefs \shortcite{VictorJeon20}. However, people often deviate from these assumptions and err in making decisions. For example, research has shown that individuals are not Bayesian in laboratory experiments and they do not always choose the payoff-maximizing options (e.g., \shortciteA{KT72, JindalAribarg21}). These errors exist even in non-social settings and for very simple decision tasks \shortcite{Benjamin18}. In social environments, decision-making is often more complex because information is hidden in others' choices and one needs to make inference about others' evaluations. Hence, decision-making errors might be inevitable in social settings. Despite this seemingly clear connection, little is known about the similarities and differences between decision-making errors in the presence and in the absence of social interaction. 

In this paper, I employ a set of simple laboratory experiments to uncover errors in learning from others' choices and contrast them with errors that are prevalent in the absence of the social element. In the experiment, subjects need to guess about an \textit{ex ante} unknown state of the world and are paid for accuracy. The state is binary and its possible realizations are represented by two boxes that contain ten balls of black and white color. The combination of black and white balls can be different across the two boxes and the content of each box is known to subjects. 
The true state is randomly realized in the beginning of the experiment, i.e., one box is selected by flipping a fair coin. Subjects do not observe the true state, but they receive a signal about it. They then guess which box is the true state. The key manipulation of the experimental design is the source of the signals: across conditions, subjects receive informationally-equivalent signals that vary only in whether or not the signal arises from a social interaction. 
In the \textit{Individual} condition (control), a subject observes a ball randomly drawn from the true state. In the \textit{Social} condition (treatment), the subject does not directly observe a ball, but she observes the choice of another participant, called \textit{neighbor} in the experiment, who has observed a ball randomly drawn from the true state. Subjects know the precise signal-generating process and that all participants are incentivized to make a correct choice. So, the provided signal is informationally equivalent across the two conditions: under common knowledge rationality, the choices of the subjects should be identical in the two conditions.

In order to identify individual errors that are associated with the social interaction (SI), I use a within-subject design. I compare subjects' choices across the individual and the social conditions and present clean evidence that, despite extensive instructions, subjects exhibit on average a higher level of irrationality in the social condition (in the presence of SI) than in the individual condition (in the absence of SI) when they receive informationally equivalent signals across the two conditions. That is, they neglect the provided information relatively more in the social condition than in the individual condition.\footnote{Throughout, I use "irrationality" and "neglect" interchangeably.} The within subject design plays a critical role here, because it isolates the errors that are associated with SI (e.g., the violation of rational expectations) while controlling for any other error that is independent of SI, e.g., errors in statistical reasoning \shortcite{Benjamin18}.\footnote{A notable difference between my work and the belief updating literature is that I examine individuals' sub-optimal "choice" under uncertainty, while this literature examines how individuals form their posterior "belief". The papers in this literature either do not collect data on actual choices or ignore the fact that having a Bayesian (non-Bayesian) belief is not necessarily equivalent to making a correct (incorrect) choice. As a result, the notion of "bias" in the belief updating literature is completely different than the notion of "error" in my study. A \textit{biased belief} is defined as a belief that is not Bayesian. However, an \textit{error in choice} is defined as an action that fails to optimize the individual's payoff based on the available information. It can be shown that neither does a biased belief necessarily lead to an error in the choice, nor is an error in the choice necessarily a result of a bias in the belief.}

A plausible explanation for the unexpected tendency to neglect information in the SI is the subject's inability to infer the relationship between what other participants choose and what they know, i.e., non-rational expectations. That is, subjects might not be able to predict how their neighbors make decision based on their private information. 
If this is the reason behind the extra neglect in the SI, then providing additional information about neighbors' behavior may help subjects to better extract the information contained in their neighbors' choices. 


To test the role of uncertainty about neighbor's behavior in the subject's irrationality, I exogenously manipulate subjects' knowledge about their neighbors across treatments by providing additional information about the neighbors. First, I design a treatment in which the neighbor is replaced by a "computer bot", whose behavior is clearly described to the subjects. The idea here is to create a social environment where there is less uncertainty in the neighbor's behavior than the baseline experiment. In this treatment, subjects are told that their neighbor is a computer bot that chooses the box with more black balls when it observes a black ball, and chooses a box with more white balls when it observes a white ball. The results of this low uncertainty treatment show that the neglect in the social condition significantly drops compared to the baseline experiment. This finding highlights that the uncertainty of neighbor's behavior can play an important role in decision-making under observational learning. In fact, the violation of rational expectations might be a result of the ambiguity in other decision makers' behavior.

Second, I devise a treatment where the subject observes both her neighbor's choice and the ball that her neighbor has seen. If the additional neglect in the SI was largely about the uncertainty in the neighbor's behavior, then the observed difference between the irrationality across the social and the individual condition should disappear in this treatment. The results support this prediction: when subjects are provided with both their neighbors' choices and the signals behind those choices, there is no statistically significant difference between the level of irrationality across the social and the individual condition. This suggests that the failure of rational expectations in the social condition is mainly driven by the ambiguity of other people's behavior, i.e., subjects behave as if they lack knowledge about how others make choice based on their private information. 

Researchers are increasingly interested in the mechanisms behind reduced-form errors in decision-making due to the view that this may help develop new behavioral models that can explain real world behavior. In the final part of the paper, I build upon the earlier reduced-form results and develop a model to identify the sources of irrationality in  decision-making under social interactions. I argue that in the context of the current study, the individual's decision-making process includes two stages. In the first stage, the subject updates her belief about the environment based on the available information. Then, in a second stage, she makes a choice based on her updated belief. I show that it is not possible to separately identify errors that occur in these two stages from choice data and one needs additional sources of variation to be able to distinguish between them. I then add a survey question to each choice that subjects make during the experiment. The purpose of this survey question is to measure the "relative" direction of the subject's posterior belief.\footnote{See \shortciteA{Manski04} for a detailed discussion of how survey data on probabilistic expectations can enable experimental economists to overcome identification problems.} Specifically, it asks about the probability that the subject believes her choice is correct. This survey question along with the subject's actual choice allow me to separate the first-stage errors from the second-stage errors. 

I show that the experimental design of this paper enables the researcher to identify three sources of error in decision-making under social interaction. Two of these sources are related to the first stage of the decision-making process and the third source is associated with the second stage. I non-parametrically estimate the two stage decision-making process and shed light on what variation in the data identifies which source of error.

The remainder of the paper proceeds as follows. Section \ref{literature} reviews the related literature. Section \ref{ex_design} describes the experimental design. Section \ref{results} presents the main results of the paper. In Section \ref{sources}, I develop a model to explain the sources of irrational choices in individual behavior and identify the channel that is influenced by SI. In section \ref{estimation}, I non-parametrically estimate a two-stage decision making process and highlight what variation in the data identifies which error in decision-making. Finally, Section \ref{conclusion} concludes the paper.

\section{Related Literature}
\label{literature}

Empirical models of choice data often assume decision makers are Bayesian (e.g., \shortciteA{ErdemKeane96}). In social environments and games of incomplete information, the assumption of rational expectations is further imposed for identification of beliefs \shortcite{VictorErhao21}. However, there are a lot of studies in the literature that document the opposite: that individuals do not always behave as a Bayesian rational agent would do. 

Plausible explanations of the violation of Bayesian rational behavior include three categories of errors: \textbf{A) \textit{belief updating error}}: this error has been widely documented in isolated environments and is not necessarily related to the social environment. It can arise due to several reasons such as \textit{conservatism} \shortcite{Edwards68,HuckWeizsacker02}, \textit{overconfidence} \shortcite{NothWeber03}, \textit{base rate fallacy} \shortcite{Goereeetal07}, or more broadly any form of non-Bayesian updating \shortcite{Grether80,JindalAribarg21,Chingetal21}. \textbf{B) \textit{reasoning error}}: this type of error often happens randomly when a subject makes decision based on her updated belief, and again is not specific to the social environment. The well-known logistic choice function is a special case of this where errors are due to random shocks that follow an extreme-value type I distribution. \textbf{C) \textit{violation of rational expectations}}: this error is specifically associated with the social environment and can be due to wrong beliefs about others' strategies \shortcite{KublerWei04,Weizsacker10} or mistrust in others' evaluations.\footnote{Some papers suggest a combination of these errors, see for example \shortciteA{MarchZiegelmeyer18,Angrisanietal20,DeFilippisetal21}.} 

Despite numerous studies in the literature, it remains unclear whether and how one can non-parametrically identify these three types of errors using empirical data. Many studies either ignore some of these errors or put additional assumptions to overcome the identification challenges. For example, the need to separate the inability to best respond to one’s beliefs (type B error) from potentially wrong beliefs about others (type C error) has been recognized in prior literature \shortcite{Stahl95,CH04}. However, most of these papers approach the identification problem by imposing parametric functional forms on beliefs and/or choice probabilities \shortcite{KublerWei04,Goereeetal07,MarchZiegelmeyer18}. These approaches might suffer from misspecification and generate misleading results \shortcite{DominitzHung09}. Some papers such as \shortciteA{Weizsacker10} utilize a reduced-form approach and do not impose parametric assumptions, but these studies are not able to precisely identify the contribution of three types of errors mentioned earlier to the subject's behavior. Hence, they provide limited explanation about the nature of the suboptimality.

The current paper contributes to the literature by developing a social interaction experiment which non-parametrically identifies these errors in subject's behavior. The framework in my study is built on the theoretical work of \shortciteA{Walliser89} and sheds light on what variation in the data identifies which error type. The idea is that type A and type C errors often appear when a subject forms expectation about the environment (stage 1 in decision-making), but type B error emerges when these expectations are used to find out a selected action (stage 2 in decision-making). To distinguish between these two stages, I collect both choice data and belief data so that I am able to separate type B errors from type A and type C errors. This highlights the fact that one cannot separately identify errors associated with the first stage of decision-making process from those of the second stage using only the choice data or only the belief data. 

To further disentangle between type A and type C errors, I separately ask subjects to process a private signal and the choice of a neighbor across two conditions with a within-subject design. The separation and within-subject design are important because without them, one may not be able to study the "absolute" effect of each source of information. In fact, as \shortciteA{Eyster19} points out, many studies in the social learning literature are only able to study the "relative" importance of private versus social information.\footnote{My study has two other advantages compared to social learning experiments \shortcite{AndersonHolt97,KublerWei04,Goereeetal07}. First, the decision task in my experiment is very simple so that there is no concern about the complexity of the decision in the social condition. The complexity of the decision problem has been shown to be an important factor in driving individual errors \shortcite{CharnessLevin09, Weizsacker10, EnkeZimmerman}. Second, by separating the people whose choices are used in the social condition from those of the main experiment, my experimental design rules out any confounding factor which might be related to the prosocial/strategic incentives among subjects and provides a lower bound on the amount of irrationality one might expect to observe in social interactions.}

While the above features of the experiment help to separate different types of error, they are not sufficient for the non-parametric identification of beliefs and choices. One needs enough variation in the information structures (precision of signals) so that beliefs cover the whole range of probabilities in $[0,1]$. The wide variety of information structures in my experiment further facilitates the non-parametric estimation and provides insights about what parametric assumptions might or might not be appropriate for modeling decisions in social interactions. This variation in the data is closely related to the literature on non-parametric identification of non-equilibrium beliefs in games of incomplete information \shortcite{VictorArvind20,VictorErhao21}. The novelty is in the source of variation which has not been sufficiently explored in prior literature. In addition, I show that there is substantial heterogeneity in decision-making across subjects. So, one needs to naturally account for individual-specific heterogeneity when modeling decision-making in social interactions. The heterogeneity in beliefs has been documented in other contexts, see for instance \shortciteA{Orhun12}, \shortciteA{Chingetal21}, \shortciteA{JindalAribarg21}, \shortciteA{BenettonCompiani21}, among others. My results are consistent with these findings and extends them to a more general framework in which there is no strategic incentive among subjects.\footnote{For a literature review on (the failure of) rational expectations in other strategic settings see \shortciteA{BeardBeil94} and \shortciteA{Eyster19}.}

\section{Experimental Design}
\label{ex_design}
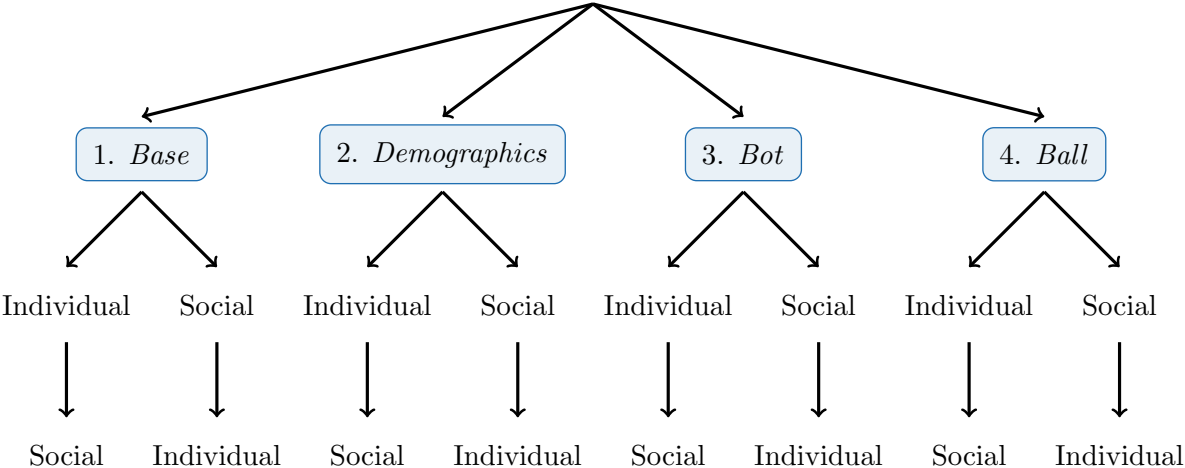
\begin{figure}[htbp]
\begin{center}
\begin{tikzpicture}[xscale=1]
\draw[->][draw=black, very thick]  (4,-1)  -- (-2,-2.5);
\draw[->][draw=black, very thick]  (4,-1) -- (2,-2.5);
\draw[->][draw=black, very thick]  (4,-1) -- (6,-2.5);
\draw[->][draw=black, very thick]  (4,-1) -- (10,-2.5);

\node[align=left] at (-2,-3)
{\mybox{1. \textit{Base}}};
\node[align=left] at (2,-3)
{\mybox{2. \textit{Demographics}}};
\node[align=right] at (6,-3)
{\mybox{3. \textit{Bot}}};
\node[align=right] at (10,-3)
{\mybox{4. \textit{Ball}}};

\draw[->][draw=black, very thick]  (-2,-3.5) -- (-3,-4.5);
\draw[->][draw=black, very thick]  (-2,-3.5) -- (-1,-4.5);

\draw[->][draw=black, very thick]  (2,-3.5) -- (1,-4.5);
\draw[->][draw=black, very thick]  (2,-3.5) -- (3,-4.5);

\draw[->][draw=black, very thick]  (6,-3.5) -- (5,-4.5);
\draw[->][draw=black, very thick]  (6,-3.5) -- (7,-4.5);

\draw[->][draw=black, very thick]  (10,-3.5) -- (9,-4.5);
\draw[->][draw=black, very thick]  (10,-3.5) -- (11,-4.5);

\node[align=left] at (-3,-5)
{Individual};
\draw[->][draw=black, very thick]  (-3,-5.5) -- (-3,-6.5);
\node[align=left] at (-3,-7)
{Social};

\node[align=left] at (-1,-5)
{Social};
\draw[->][draw=black, very thick]  (-1,-5.5) -- (-1,-6.5);
\node[align=left] at (-1,-7)
{Individual};

\node[align=left] at (1,-5)
{Individual};
\draw[->][draw=black, very thick]  (1,-5.5) -- (1,-6.5);
\node[align=left] at (1,-7)
{Social};

\node[align=left] at (3,-5)
{Social};
\draw[->][draw=black, very thick]  (3,-5.5) -- (3,-6.5);
\node[align=left] at (3,-7)
{Individual};

\node[align=left] at (5,-5)
{Individual};
\draw[->][draw=black, very thick]  (5,-5.5) -- (5,-6.5);
\node[align=left] at (5,-7)
{Social};

\node[align=left] at (7,-5)
{Social};
\draw[->][draw=black, very thick]  (7,-5.5) -- (7,-6.5);
\node[align=left] at (7,-7)
{Individual};

\node[align=left] at (9,-5)
{Individual};
\draw[->][draw=black, very thick]  (9,-5.5) -- (9,-6.5);
\node[align=left] at (9,-7)
{Social};

\node[align=left] at (11,-5)
{Social};
\draw[->][draw=black, very thick]  (11,-5.5) -- (11,-6.5);
\node[align=left] at (11,-7)
{Individual};

\end{tikzpicture}
\end{center}
\caption{The experimental design}
\label{design}
\end{figure}

Subjects are randomly assigned to one of four treatments (Figure \ref{design}). The experiment in each treatment consists of two consecutive parts and the order of these two parts is randomized. In one part, the subject performs a task in isolation --- without social interaction (\textit{individual condition}). In the other part, she performs the same task with social interaction (\textit{social condition}). As noted earlier, the order of these two parts is randomized, i.e., given a treatment, some subjects first see the individual condition and then proceed to the social condition, and others see them in reverse order (see Figure \ref{design}). 
\\

The individual condition is the same in all four treatments, but the social condition differs across treatments. The idea is to exogenously manipulate the subject's knowledge about the participants with whom she is interacting across treatments. I will elaborate on the differences between treatments as I proceed in the following. 

\subsection{Individual Condition}
\begin{figure}[htbp]
\centering
        \includegraphics[scale=0.6]{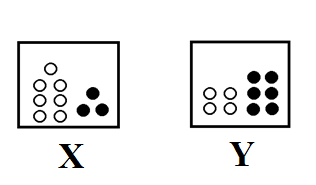}
      \caption{These two boxes represent possible states of the world (Here, for instance $\theta_X=0.7$ and $\theta_Y=0.6$)}
      \label{URNS}
\end{figure}

The \textit{individual condition} is a benchmark which measures the subjects' behavior in the absence of social interaction. It consists of 21 rounds. In each round, two boxes are shown to the subject. Each box contains 10 balls of white or black color (see Figure \ref{URNS} for an example). These boxes represent the possible states of the world, $\omega \in \{X,Y\}$. In the beginning of each round, a fair coin is anonymously flipped. If the coin is \textbf{Head}, the state is \textbf{X} and one ball is randomly drawn from box X. If the coin is \textbf{Tail}, the state is \textbf{Y} and one ball is randomly drawn from box Y.\footnote{This induces a prior probability of $\frac{1}{2}$ for each box. The language used in the actual experiment was slightly different: I used \textit{box }$H$ (head) and \textit{box }$T$ (tail) instead of box $X$ and box $Y$ to remind individuals about the randomization (see the appendix for experiment instructions).} The subject does not observe the coin. She observes the ball, and then is asked to guess what the state is. The combination of white and black balls randomly changes over 21 rounds. Denote the fraction of white balls in box $X$ by $\theta_X$ and the fraction of black balls in box $Y$ by $\theta_Y$. The combinations used in the experiment include a wide range of symmetric and asymmetric information structures: $\Big\{(\theta_X,\theta_Y) \ \Big| \ \theta_X,\theta_Y \in \{0.5,0.6,0.7,0.8,0.9,1\} \ , \ \theta_X \geq \theta_Y \Big\}$.
\\
Subjects are incentivized to make a correct guess:\footnote{The idea is to randomly choose some rounds and pay the subject for each correct guess in those rounds. I explain the payment scheme in more details later.} it is best for a subject to pick a box with more black (white) balls when she observes a black (white) ball. In addition to collecting the subjects' choices, I add a survey question at the end of each round that asks for the subject's posterior belief. Specifically, the subject answers the following question after she reports her choice in each round: \textit{with what probability do you think your guess is correct?}.\footnote{As I explain later, I am interested in knowing which state is more likely from the subject's perspective when she makes a choice. Effectively, I only need to know whether the subject chooses a state that she believes has a higher ($>$50\%) or lower ($<$50\%) chance of being correct.} Unlike the choice, the survey question is not incentivized in the experiment for a few reasons. First, I found that the experiment lasts too long when subjects are required to go through an incentive compatible elicitation procedure for each of the posterior beliefs that they submit during the experiment. So, it may cause fatigue and contaminate the choice data that is vital for the main analysis. Second, I did use monetary incentives for posteriors in a pilot study using a revised version of Quadratic Scoring Rule \shortcite{Brier50}.\footnote{For the subjects who are incentivized for both the choice and the posterior, I randomly select one posterior and one choice for payment (the selection is independent).} The pilot results suggested that the incentivized posteriors are not significantly different than the posteriors that are not incentivized. The prior literature has also shown that responses to this type of survey questions, in the absence of incentives for honest revelation of expectations, do possess face validity when the questions concern well-defined events; see \shortciteA{Manski04} for a detailed discussion. 
I kept the survey question very standard and easy to understand so that it is unlikely that the subject does not understand the survey question or incurs a cognitive cost to think about the answer \shortcite{Smith91}. So, one can expect the reported posterior probabilities to be close to the subjective probabilities in the subject's mind.\footnote{I also did a robustness check at the end of my main experiment and incentivized all subjects according to Quadratic Scoring Rule. The elicited posteriors were very similar to those that were collected from the survey questions during the experiment. But I do not use these incentivized posteriors in my analysis because the incentive compatible elicitation of posteriors were always happening at the end of the experiment, after both the individual condition and the social condition had been finished. The fact that the incentivized posteriors were always collected after the end of the experiment might make the results inconclusive (the subjects were answering the same survey questions as they had observed during the experiment. The concern is that subjects might not think about the questions anymore because they had already seen the same questions before. Hence, elicitation mechanism might not have an impact on subjects' posteriors.)} 
 
\subsection{Social Condition}
The \textit{social condition} is designed to study the subjects' behavior in the presence of social interaction. The structure of the task is similar to that of the individual condition. The social condition consists of 21 rounds. In each round, the subject is randomly connected to another participant, called \textit{neighbor} in the experiment, and receives information from one of the rounds in the neighbor's individual condition. The subject observes the content of two boxes that has been shown to the neighbor. Her task is to guess what the state (selected box) is, based on the information that she obtains from the neighbor. As noted earlier, there are four treatments in the experiment and the transmitted information in the social condition is different across treatments. I explain the treatments in the following.

The first treatment is called \textit{base}. In this treatment, the information coming through social interaction is the neighbor's guess.\footnote{By "guess" I mean the actual choice of the neighbor. The subject does not observe the neighbor's posterior belief (the answer to the survey question).} Note that a neighbor here is a random subject who has previously participated in the experiment. The subject knows that the neighbor's guess is incentivized and is based on a randomly drawn ball from the realized state (box). To summarize, in each round, the subject observes two boxes and the guess of a neighbor, but not the ball that the neighbor has observed. Then, the subject is asked to guess about the realized state. The experiment is designed such that the neighbor randomly changes in each round. Hence, the subject does not interact with the same neighbor over time and it is unlikely that the subject learns about a specific neighbor's behavior over the course of 21 rounds in the social condition.

The second treatment is called \textit{demographics}. There is a slight difference between the social condition in this treatment and in the base treatment: on top of the neighbor's guess, the subject observes the neighbor's demographic information such as age, gender, years of education, and whether the neighbor has taken any Probability/Statistics courses. This treatment is designed to examine whether providing demographic information about the neighbor can alleviate the irrationality associated with the uncertainty of neighbor's behavior in the social interaction. If demographics provide additional information about the behavior of neighbor, the uncertainty might be lower in this treatment than the base treatment.

The third treatment is called \textit{bot}. Everything in this treatment is the same as in the base treatment, except that the neighbor is a computer bot which is programmed to exhibit a specific behavior (i.e., rational). This means when the bot observes a white ball, it chooses the box with more white balls, and when it observes a black ball, it chooses the box with more black balls. The behavior of the bot is explained in details to the subjects in this treatment. Subjects see the guess of the bot in this treatment and then submit their own guesses about the realized state. The social interaction in this treatment is relatively more transparent than the earlier two treatments. So, one expects the irrationality associated with the uncertainty of neighbor's behavior to be significantly lower in this treatment than the base treatment.

The fourth treatment, which is called \textit{ball}, is an augmented version of the base treatment in which the subject observes both the (human) neighbor's guess and the ball which has been shown to the neighbor. The uncertainty effect is expected to completely disappear in this treatment because the subject is provided with all the relevant information regarding her neighbor's choice.

\subsection{Payment Scheme}
Each subject receives \$6 show-up fee for participation. In addition to that, two rounds of the experiment are randomly selected and the subject wins \$12 for each correct guess in those two rounds.

\section{Results}
\label{results}
In this section, I first define the criteria for recognizing individual errors in the context of my experiment. I then analyse subjects' choices in the experiment to measure the frequency of these errors and examine the relation between them in the individual condition and in the social condition. The comparison between errors in the individual and in the social condition isolates the errors that are independent of the social interaction (belief updating error and reasoning error) and identifies the errors that are associated with the social interaction (violation of rational expectations).

In the individual condition, a Bayesian rational subject should choose a box with more black balls when she observes a black ball, and a box with more white balls when she observes a white ball. Accordingly, I define an \textit{individual irrationality} as an observation which deviates from this prediction.

\begin{definition}
Individual Irrationality: A choice in the individual condition where the subject observes a white (black) ball, but chooses a box with more black (white) balls.
\end{definition}

In social interactions, the conventional assumption in economics is that individuals have rational expectations about each other (and rationality is common knowledge). In the context of the current experiment, this implies that the subject should follow her neighbor's guess and choose the same box as the neighbor in the social condition. Accordingly, a \textit{social irrationality} is defined as follows.

\begin{definition}
Social Irrationality: A choice in the social condition where the subject chooses a box different from her neighbor's guess.\footnote{The definition of social irrationality in the \textit{ball} treatment is a little bit different because the subject observes both the ball and the neighbor's guess when she is connected to the neighbor. In that case, I define social irrationality as a choice in which the subject chooses a box different from her neighbor's guess, given that the neighbor's guess is rational (i.e., does not contradict with the signal).}
\end{definition}

In the next section, I analyse the experimental data to measure the magnitude of individual irrationality and social irrationality in subjects' choices and to elaborate on the differences.

\subsection{Data}
The main experiment was conducted 
at Toronto Experimental Economics Laboratory (TEEL) in University of Toronto 
during December 2019. The experiment was programmed in oTree \shortcite{oTree}. In total, 151 subjects were recruited from the subject pool using Online Recruitment System for Economic Experiments \shortcite{Greiner15}. The average payment across subjects was \$25.26.\footnote{No subject participated in more than a single treatment. Subjects needed to be at least 18 years old to be eligible to participate in the experiment. The human neighbors in the social condition were 94 subjects who had participated in the experiment a few months before the main experiment.} 
Table \ref{Summary_stat} provides evidence that individual characteristics are relatively balanced across the four
treatments, confirming that the randomization was successful.
\begin{table}[ht] \centering 
  \caption{Summary Statistics} 
  \label{Summary_stat} 
\begin{tabular}{@{\extracolsep{5pt}}lcccc} 
\\[-1.8ex]\hline 
\hline \\[-1.8ex] 
 & \multicolumn{4}{c}{Treatment} \\ 
\cline{2-5} 
\\[-1.8ex] & Base & Demographics & Bot & Ball\\
\hline \\[-1.8ex] 
 Female (\%) & 77.5 & 73.5 & 78.9 & 61.5 \\ 
  & & & & \\ 
 Prob/Stat course (\%) & 75 & 64.7 & 68.4 & 69.2 \\ 
  & & & & \\ 
 Years of Education & 15.0 & 14.85 & 14.73 & 14.48 \\ 
  & (1.5) & (2.11) & (2.24) & (1.82) \\ 
  & & & & \\ 
 Age & 20.25 & 19.38 & 20.02 & 20.28 \\ 
  & (1.81) & (1.39) & (2.04) & (1.88) \\ 
  \hline \\[-1.8ex] 
Number of Subjects & 40 & 34 & 38 & 39 \\ 
\hline 
\hline \\[-1.8ex]
\end{tabular} \\
      \small \textit{Note:} Standard deviations are presented in parentheses. The second row shows \\ the percentage of subjects who have taken Probability/Statistics courses.
\end{table} 

In the following, I exclude the cases in which both boxes have 5 black balls and 5 white balls, $\theta_X=\theta_Y=0.5$, because theory does not have a prediction about the subject's behavior in those cases. Subjects are expected to behave randomly in those rounds, a result that is supported by the data.\footnote{In the individual condition, when the two boxes have the same combination of balls (5 white and 5 black balls), subjects choose the left box with probability 0.44. Here, the null $H_0:p = 0.5$ cannot be rejected at the 5\% significance level (\textit{p\text{-}value} = 0.14). Similarly, in the social condition, when both boxes have 5 white and 5 black balls, subjects do not follow their neighbor's guess with probability 0.48 ($p\text{-}value=0.74$ for the null $H_0:p=0.5$).}

\subsection{How Do Errors Differ across the Individual and the Social Conditions?}

\begin{figure}[htbp]
\centering
        \includegraphics[scale=0.6]{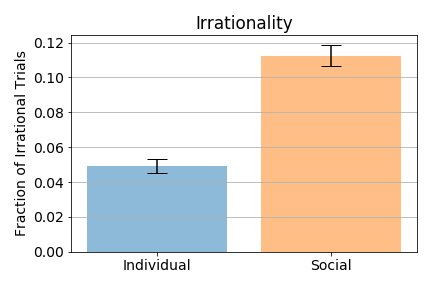}
      \caption{Individual irrationality and social irrationality (pooled data)}
      \label{Ind-vs-Soc-total-within_corrected}
\end{figure}

My first result examines the aggregate fraction of irrational choices in the individual condition and in the social condition. Figure \ref{Ind-vs-Soc-total-within_corrected} illustrates that the individual irrationality and the social irrationality are significantly greater than zero, even though subjects are incentivized for being correct. In the individual condition, subjects on average deviate from the theoretical prediction (Bayesian rational behavior) with a probability of 0.049 ($p\text{-}value<0.001$).\footnote{This is lower than the error rate reported in prior literature for the individuals who arrive first in a standard social learning experiment \shortcite{AndersonHolt97,KublerWei04}. In \shortciteA{AndersonHolt97} 10\% of the subjects whose information set was a private signal, did not follow their signal. In \shortciteA{KublerWei04}, this behavior was observed in about 7\% of all cases where first players saw only a private signal.} In the social condition, even though subjects know that their neighbor's guess is incentivized with money, they on average do not follow the choice of their neighbor with a probability of 0.112 ($p\text{-}value<0.001$). Surprisingly, the social irrationality is significantly higher than the individual irrationality ($p\text{-}value<0.001$). This evidence suggests that subjects neglect the information more in the social condition than in the individual condition, a result that can be associated with, for example, the violation of rational expectations.\footnote{Note that the comparisons in this section are within-subject, i.e., the same subjects are making on average more errors in the social condition than in the individual condition. Given my experimental design, it is also possible to do the analysis between-subject. The details of the between-subject analysis are provided in the appendix. The results are qualitatively similar there.}

\begin{figure}[htbp]
\centering
        \includegraphics[scale=0.75]{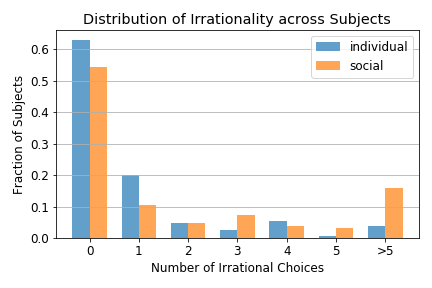}
      \caption{The distribution of individual irrationality and social irrationality across subjects}
      \label{irr_hist}
\end{figure}

Figure \ref{irr_hist} presents the distribution of individual irrationality and social irrationality across subjects. The blue histogram shows that about 63\% of subjects have no individual irrationality over the course of 21 rounds in the individual condition. In addition, 19.8\% of subjects have exactly one individual irrationality, and the remaining 17.2\% have more than one individual irrationality. So, the individual irrationality is not negligible for a considerable fraction of subjects.\footnote{This result is consistent with \shortciteA{AmbuelLi18} who report that 17\% of their subjects made at least one irrational choice out of six trials.} On the other hand, the orange histogram indicates that 54.3\% of subjects have no social irrationality, 10.6\% have exactly one social irrationality, and the remaining 35.1\% have more than one. Comparing the two distributions, one can observe that the upper tail of the distribution is thicker in the social condition than in the individual condition. So, there is a clear shift in the error rate of subjects across the two conditions. The two-sample Anderson-Darling test also verifies the significant difference between the two distributions ($p\text{-}value<0.005$).

\subsection{The Role of Uncertainty about Others' Behavior}

\begin{figure}[htbp]
\centering
        \includegraphics[scale=0.6]{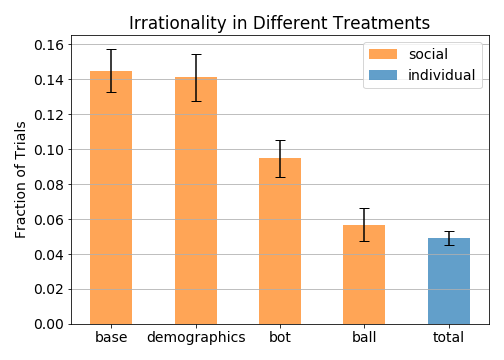}
      \caption{Social irrationality in different treatments}
      \label{Ind-vs-Soc-treatments_corrected}
\end{figure}

In this section, I examine the mechanisms behind the additional neglect in the social condition. The hypothesis is that the additional irrationality in the social condition arises from the uncertainty about the neighbor's behavior. In other words, because the neighbor's decision-making process is ambiguous to the subjects, they cannot correctly extract the neighbor's private information from observation of their choices, and thus violate the rational expectations. To test this idea, as noted earlier, I exogenously manipulate the subject's knowledge about the neighbor across four treatments. Figure \ref{Ind-vs-Soc-treatments_corrected} illustrates the social irrationality in each treatment along with the aggregate individual irrationality.\footnote{Recall that the individual condition is identical in all treatments. So, I do not break down the individual irrationality here and only report the aggregate individual irrationality for ease of exposition.} The treatment "\textit{base}" is a benchmark treatment in which subjects only observe the guess of their neighbor. The result in this treatment echos the earlier finding about the larger magnitude of neglect (irrationality) in the social condition than in the individual condition. 

In the treatment "\textit{demographics}", subjects are provided with some demographic information about their neighbor (Age, Gender, Years of Education, Whether the neighbor took Probability/Statistics courses), on top of the neighbor's guess. The result of this treatment shows that providing demographics slightly decreases the social irrationality compared to the base treatment, from $0.145$ to $0.1411$. But this effect is not statistically significant ($p\text{-}value=0.83$).

In the treatment "\textit{bot}", subjects observe the guess of a computer bot. Here, the bot's behavior is known to subjects: it picks a box with more white balls when it observes a white ball, and picks a box with more black balls when it observes a black ball. The social irrationality in this treatment significantly drops to 0.094 compared to the base treatment ($p\text{-}value<0.01$). This evidence is consistent with the hypothesis that the difference between the social irrationality and the individual irrationality is due to the uncertainty about the neighbor's behavior.\footnote{Note that although the social irrationality is alleviated in the bot treatment, it is still significantly higher than the individual irrationality. One natural question rises here: why is there a difference between the individual irrationality and the social irrationality in the bot treatment? Responses from an open survey question that was collected at the end of the experiment show that some of the subjects mistrust bots. This might explain why the social irrationality in the bot treatment remains significantly higher than the individual irrationality.}

Finally, in the treatment "\textit{ball}", subjects are provided with both the neighbor's guess and the ball which was shown to the neighbor. Figure \ref{Ind-vs-Soc-treatments_corrected} shows that the social irrationality in this treatment is significantly lower than all other treatments ($p\text{-}value<0.01$). Here, the difference between the social irrationality and the individual irrationality is no longer statistically significant. This result verifies that when there is no uncertainty about the neighbor's behavior in the social condition, the magnitude of social irrationality is the same as the magnitude of individual irrationality. So, the additional neglect in the social condition disappears from the subject behavior.

\subsection{The Observed Heterogeneity in Subjects' Behavior}
In this section, I run some regressions to examine the observed heterogeneity in the subjects' behavior. My data contains demographic information about all the subjects and each of their neighbors. So, I can investigate how subjects' characteristics and those of their neighbors explain the observed irrationality in the experiment. 

First, I investigate the role of the subject's characteristics. Specifically, I estimate the following regression,
\begin{align}
    Y_{i}\times 100 =  \alpha+X_i\gamma+\epsilon
\end{align}
where $Y_{i}$ is the fraction of irrational choices by subject $i$, $X_i$ includes the subject's observed characteristics: gender (dummy for female), years of education, age, and whether the subject has taken Probability/Statistics courses. The estimation results are provided in Table \ref{Regress}. 
\\

\begin{table}[ht] \centering 
  \caption{The observed heterogeneity in the subject's irrationality} 
  \label{Regress} 
\begin{tabular}{@{\extracolsep{5pt}}lcc} 
\\[-1.8ex]\hline 
\hline \\[-1.8ex] 
 & \multicolumn{2}{c}{Data} \\ 
\cline{2-3} 
\\[-1.8ex] & Individual & Social\\
\\[-1.8ex] & (1) & (2)\\
\\[-1.8ex] DV & $Y_{ic}\times 100$ & $Y_{ic}\times 100$\\
\hline \\[-1.8ex] 
 Gender (Female) & 0.47 & $5.58^*$ \\ 
  & (1.84) & (2.88) \\ 
  & & \\ 
 Education & 1.00$^{**}$ & 0.11 \\ 
  & (0.49) & (0.78) \\ 
  & &  \\ 
 Age & $-$ 0.51 & $-$ 1.36 \\ 
  & (0.54) & (0.85) \\ 
  & &\\
 Prob/Stat course & $-$ 4.02$^{**}$ & $-$ 6.85$^{**}$ \\ 
  & (1.89) & (2.95) \\
  & & \\
  Constant &2.81 & 37.19$^{**}$ \\ 
  & (9.35) & (14.64) \\ 
  \hline \\[-1.8ex] 
Observations  & 151 & 151 \\ 
R$^{2}$  & 0.063 & 0.097 \\ 
Adjusted R$^{2}$  & 0.037 & 0.073 \\ 
\hline 
\hline \\[-1.8ex] 
\textit{Note:}  & \multicolumn{2}{r}{$^{*}p<0.1$; $^{**}p<0.05$; $^{***}p<0.01$} \\ 
\end{tabular} 
\end{table} 
%
Column (1) in Table \ref{Regress} indicates that in the individual condition, all else equal, subjects who have taken Probability/Statistics courses make 4.02 percentage points less errors than subjects who have not. This result suggests that the individual irrationality is mainly driven by the lack of knowledge in probability and statistics. Column (2) illustrates that in the social condition, all else equal, the subjects who have taken Probability/Statistics courses make 6.85 percentage points less errors than subjects who have not. My results are consistent with the previous findings in the literature. For example, in a different context, \shortciteA{Armantieretal15} find evidence that respondents whose behavior cannot be rationalized by economic theory, tend to have lower education and lower numeracy and financial literacy. In my experiment, however, the subject's observable characteristics cannot explain the additional irrationality that is associated with the uncertainty about the neighbor's behavior in the social condition (i.e., violation of rational expectations).

Next, I examine the effect of the neighbor's observable characteristics on the subject's social irrationality. Note that only the subjects who are in the treatment \textit{demographics} observe their neighbor's characteristics. So, I need to restrict the data in this section to the choices made in the social condition of the demographics treatment. Here, the dependent variable is a binary choice. Hence, I estimate the following logistic regression,
\begin{align}
\label{logistic_social}
    Pr(D_{ij} \text{ is irrational}) = \frac{\text{exp}\ (\alpha+X_i\gamma+X_j \delta)}{1+\text{exp}\ (\alpha+X_i\gamma+X_j \delta)}
\end{align}
where $D_{ij}$ is the choice of subject $i$ in round $j$, $X_i$ includes the subject's observable characteristics, and $X_j$ includes the neighbor's observable characteristics in round $j$ (recall that the neighbor randomly changes in each round). As before, observable characteristics include gender, years of education, age, and whether the individual has taken Probability/Statistics course. The estimation results are shown in Table \ref{Regress neighbor}.
%
%
%

\begin{table}[htp] \centering 
  \caption{The observed heterogeneity in the social condition of treatment "\textit{demographics}" (Equation \ref{logistic_social})} 
  \label{Regress neighbor} 
\begin{tabular}{@{\extracolsep{5pt}}lcccc} 
\\[-1.8ex]\hline 
\hline \\[-1.8ex] 
 & \multicolumn{2}{c}{Logit Coefficients} & \multicolumn{2}{c}{Average Marginal Effects} \\ 
\cline{2-3} \cline{4-5}
\\[-1.8ex] & (1) & (2) & (3) & (4) \\
\hline \\[-1.8ex] 
 Subject's Gender (Female) & 0.52 & 0.48 & 0.062 & 0.056\\ 
  & (0.59) & (0.60)  & (0.071) & (0.07)\\ 
  & & & & \\ 
 Subject's Education & $-$0.139 & $-$ 0.159 & $-$ 0.016 & $-$ 0.018 \\ 
  & (0.17) & (0.17) & (0.02) & (0.02)  \\ 
  & &  & & \\ 
 Subject's Age & $-$ 0.08 & $-$ 0.075 & $-$ 0.009 & $-$ 0.008  \\
  & (0.23) & (0.23) & (0.0267) & (0.026)\\ 
  & &  & &\\
 Subject's  Prob/Stat course & $-$ 0.44 &  $-$ 0.46  & $-$ 0.052 & $-$ 0.052\\ 
  & (0.63) & (0.64) & (0.076) & (0.076)  \\ 
  & &   & &\\ 
  Neighbor's Gender (Female) & & $-$ 0.41$^{**}$ & & $-$ 0.047$^{*}$ \\ 
  & & (0.20) & & (0.025)  \\ 
  &  &  & &\\ 
 Neighbor's Education & & $-$ 0.04  & & $-$ 0.005 \\ 
  & & (0.032) & & (0.003)\\ 
  &  &  & &\\ 
 Neighbor's Age & & $-$ 0.025$^{**}$  & & $-$ 0.003$^{***}$   \\ 
  & & (0.01) & & (0.001) \\ 
  & &  & &\\
 Neighbor's Prob/Stat course & & $-$ 0.51$^{***}$ & & $-$ 0.058$^{**}$   \\ 
  & & (0.196) & & (0.025)\\ 
  &  &  & &\\ 
 Constant & 2.18 & 4.19  & &\\ 
  & (3.2) & (3.51)& & \\ 
\hline \\[-1.8ex] 
Observations & 680 &  680 & 680 & 680\\ 
Pseudo R$^{2}$ & 0.035 & 0.061 & 0.035 & 0.061\\ 
\hline 
\hline \\[-1.8ex] 
\end{tabular} \\
\textit{Note}: Standard errors are clustered at the subject level.\\$^{*}p<0.1$; $^{**}p<0.05$; $^{***}p<0.01$
\end{table} 

Columns (1) and (2) in Table \ref{Regress neighbor} present the estimated coefficients for equation \eqref{logistic_social}. The coefficients are insignificant in the first column. However, the second column shows that the neighbor's observable characteristics have a statistically significant effect on the subject's behavior: \textit{ceteris paribus}, the subject is more likely to follow a neighbor whose age is higher, whose gender is female (versus male), and who has taken Probability/Statistics courses. 

The coefficients of a logistic regression are not quantitatively interpretable. So, I report the average marginal effects in columns (3) and (4) of Table \ref{Regress neighbor}. The results in column (4) imply that, \textit{ceteris paribus}, a subject is likely to follow a neighbor who has taken Probability/Statistics courses 5.8 percentage points more than a neighbor who has not. In addition, all else equal, a subject is 4.7 percentage points less likely to make an irrational guess when interacting with a female versus a male neighbor (4.7 percentage points more likely to follow the neighbor). The effect of the neighbor's age is very small though, i.e., one year increase in the neighbor's age increases the likelihood of being followed by the subject by 0.3 percentage points.

\subsection{Heterogeneity across Information Structures}
\begin{figure}[htbp]
\centering
    \includegraphics[scale=0.55]{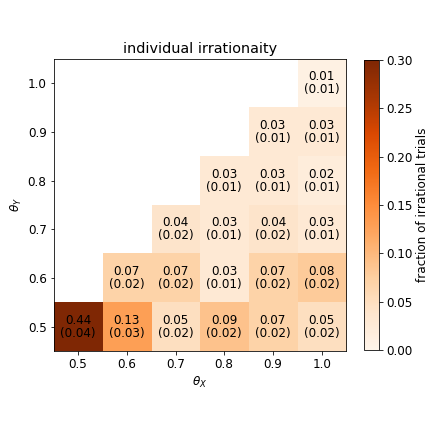}%
    \includegraphics[scale=0.55]{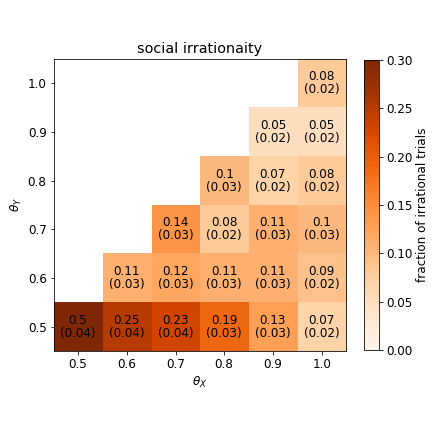}%
    \caption{The fraction of irrational choices conditional on the information structure. Standard errors are shown in parentheses.}
    \label{fig:info_struct}
\end{figure}

A unique feature of the experiment is that subjects make decisions given several information structures. This allows me to examine the extent to which subjects' irrationality vary across signal precision. Figure \ref{fig:info_struct} shows the total fractional of irrational choices conditional on each information structure. The general trend is that the irrationality decreases as signal precision increases. However, even in the extreme case of complete certainty ($\theta_X=\theta_Y=1$) the irrationality is not zero. The notable insight is that numbers are significantly higher in the social condition (right figure) than in the individual condition (left figure) which is consistent with my earlier findings.  

\begin{figure}[htbp]
\centering
    \includegraphics[scale=0.55]{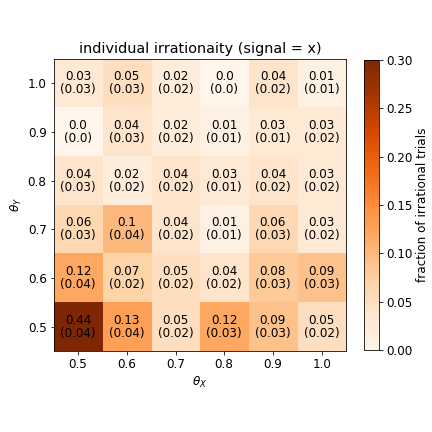}%
    \includegraphics[scale=0.55]{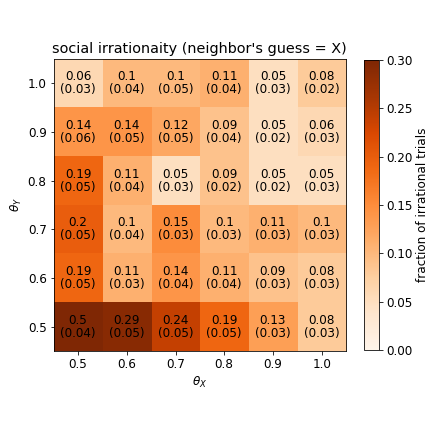}%
    \caption{The fraction of irrational choices conditional on the information structure and the signal being $x$ (in the individual condition) or a neighbor's guess of $X$ (in the social condition). Standard errors are shown in parentheses.}
    \label{fig:info_struct_sig}%
\end{figure}

Note that even though each figure covers half of the possible combinations of $\theta_X$ and $\theta_Y$, these results are equally applicable to the uncovered cases. This is because if one exchanges the contents of box X and box Y, the environment remains the same but the information structure changes from $(\theta_X,\theta_Y)$ to $(\theta_Y,\theta_X)$. Hence, one can naturally extend these figures so that they cover the whole space of information structures. 

Another way to examine subjects' behavior is to measure irrational choices conditional on both the information structure and the realized signal. This is important specially in cases where the information structure is asymmetric. As an example, consider a case in which $\theta_X=1$ and $\theta_Y=0.5$. In this case, box X has 10 white balls and box Y has 5 white and 5 black balls. Here, when the signal is a black ball, a Bayesian subject is certain that the state is box B. However, when the signal is a white ball, the Bayesian subject assigns a relatively higher likelihood to box X than box Y, but she would not be 100\% sure about the state. So, one might expect to see a higher irrationality in the latter case than the former case. 

Without loss of generality and for expositional purposes, I permute observations so that all the signals in the individual condition are a white ball, and all the signals in the social condition are a neighbor's guess of $X$. This is possible because of the nature of the experiment. For example, an observation in the individual condition where $(\theta_X,\theta_Y)=(0.5,1)$ and the signal is $y$ (black ball), can be permuted so that $(\theta_X,\theta_Y)=(1,0.5)$ and the signal is $x$ (white ball).\footnote{This way, one only needs to plot one figure for each condition, and the figure shows irrationalities conditional on all information structures and signal "$x$". If I do not permute observations, then I need to plot two figures for each condition: one figure would be conditioned on signal $x$ (but covers half of the information space as in Figure \ref{fig:info_struct}) and the other figure would be conditioned on signal $y$.} The results are presented in Figure \ref{fig:info_struct_sig}. 

Figure \ref{fig:info_struct_sig} demonstrates that errors are generally higher in cases where the uncertainty is higher. For example, when the subject observes $x$ conditional on $(\theta_X,\theta_Y) = (1,0.5)$, the irrationality is higher than the mirror case in which $(\theta_X,\theta_Y) = (0.5,1)$. As discussed, this is the case because when a subject observes signal $x$ in the former case, she is not certain about the state. However, observing the same signal in the latter case, the subjects would be almost certain about the state. Comparing the results across conditions, the numbers are generally higher in the social condition than the individual condition. So, even though subjects react to changes in the likelihood, the uncertainty about the neighbor's behavior remains stable across information structures.

\section{A Model of Decision-making} \label{sources}
So far, I compared irrational choices across the individual and the social conditions to disentangle between errors that are independent of the social environment and errors associated with the social interaction. I showed in a reduced-form sense that rational expectations can be violated in social interaction. In this section, I build on these reduced-form results and introduce a framework to describe the individual decision-making process in the context of my experiment. The purpose of this framework is to elaborate on the fundamentals of the individual behavior under social interaction and identify sources of error in decision-making. I will estimate a non-parametric model based on this framework in the next section. 

\begin{figure}[htbp]
\begin{center}
\begin{tikzpicture}[xscale=1]
\draw[->][draw=black, very thick]  (-1,0)  node [left] {Input (signal)} -- (0.5,0);;
\draw[-][draw=blue, very thick]  (0,0.75) -- (9,0.75);
\draw[-][draw=blue, very thick]  (0,-0.75) -- (9,-0.75);
\draw[->][draw=black, very thick]  (3.5,0) -- (5.5,0);

\draw[->][draw=black, very thick]  (8.5,0) -- (10,0) node [right] {Output (choice)};
\draw [draw=blue, very thick](0,-0.75)  -- (0, 0.75);
\draw [draw=blue, very thick](9,-0.75)  -- (9, 0.75);

\node[align=left, below, color=blue] at (-2,-1.1)
{$x$};
\node[align=left] at (2,0)
{Update Belief};
\node[align=left, below, color = blue] at (2.5,-1)
{$Pr(X|x)>Pr(Y|x)$};
\node[align=center, above] at (4.5,1)
{Decision-Making Process};
\node[align=right] at (7,0)
{Binary Decision};
\node[align=left, below, color = blue] at (7,-1)
{$X$};
\node[align=left, below, color = blue] at (11,-1)
{$X$};
\end{tikzpicture}
\end{center}
\caption{The individual decision-making process}
\label{decision process}
\end{figure}
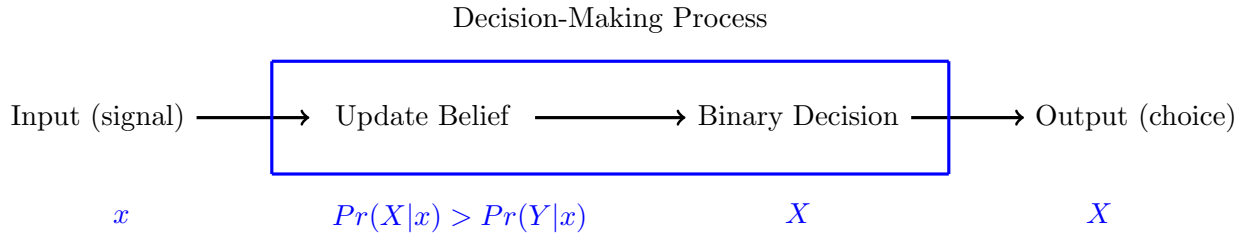

In the context of my experiment, the individual's decision-making process can be modeled as a two stage procedure. Upon observing a signal, the subject first updates her belief and then picks one of the two possible states based on her posterior. This process is shown in Figure \ref{decision process}.
\\

\begin{figure}[htbp]
\begin{center}
\begin{tikzpicture}[xscale=1]
\draw[->][draw=black, very thick]  (-1,0)  node [left] {Input (signal)} -- (0.5,0);;
\draw[-][draw=blue, very thick]  (0,0.75) -- (9,0.75);
\draw[-][draw=blue, very thick]  (0,-0.75) -- (9,-0.75);
\draw[->][draw=black, very thick]  (3.5,0) -- (5.5,0);

\draw[->][draw=black, very thick]  (8.5,0) -- (10,0) node [right] {Output (choice)};
\draw [draw=blue, very thick](0,-0.75)  -- (0, 0.75);
\draw [draw=blue, very thick](9,-0.75)  -- (9, 0.75);

\node[align=left, below, color=black] at (-2,-1)
{$x$};
\node[align=left] at (2,0)
{Update Belief};
\node[align=left, below, color = red] at (2.5,-2)
{$Pr(X|x)<Pr(Y|x)$};
\node[align=left, below, color = black] at (2.5,-3)
{$Pr(X|x)>Pr(Y|x)$};
\node[align=center, above] at (4.5,1)
{Decision-Making Process};
\node[align=right] at (7,0)
{Binary Decision};
\node[align=left, below, color = black] at (7,-2)
{$Y$};
\node[align=left, below, color = red] at (7,-3)
{\large{$Y$}};
\node[align=left, below, color = black] at (11,-1)
{$Y$};
\node[align=left, below, color = red] at (11,-2)
{posterior error};
\node[align=left, below, color = red] at (11,-3)
{reasoning error};
\end{tikzpicture}
\end{center}
\caption{Two explanations for an observed error in the individual's choice}
\label{decision error}
\end{figure}
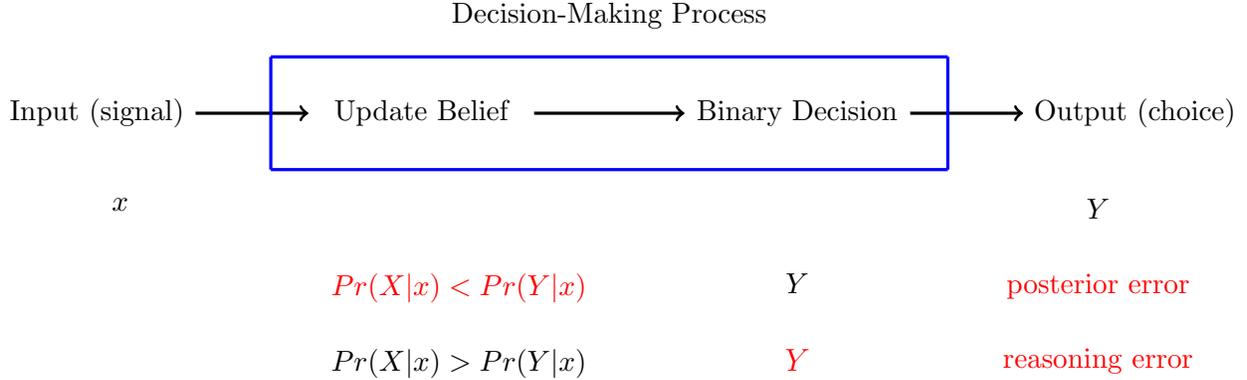

This model is closely related to the deliberation process introduced in a more general decision framework by \shortciteA{Walliser89}. In the first stage, the subject treats the available data to form expectations on her environment ("cognitive rationality"). In the second stage, these expectations are used to find out a selected action ("instrumental rationality"). As an example, suppose an individual observes signal $x$. In theory, the individual should update her belief in favor of state $X$ in the first stage, $Pr(X|x)>Pr(Y|x)$, and then choose state $X$ in the second stage. However, as I showed earlier, the individual's choice (output) does not always comply with the signal (input). Individuals frequently make errors in this simple task. Now, consider a case in which a subject obtains signal $x$, but her final guess is $Y$ (Figure \ref{decision error}). There are two explanations for this observation: 
\begin{enumerate}
\item \textbf{Posterior Error}: It might be that the subject's posterior is mistakenly in favor of state $Y$, $Pr(Y|x)>Pr(X|x)$, and this causes the subject to make an erroneous decision. This means that the subject's posterior is in a wrong direction, but her choice is consistent with her incorrect posterior.\footnote{It is important to notice that the posterior error is different from what is commonly known as "belief updating bias" in the literature \shortcite{KT72,Benjamin18}. A biased belief is not necessarily in a wrong direction, i.e., the biased belief and the Bayesian belief can both favor the same state while assigning different likelihoods to that state. For example, when the signal implies a 70\% chance (in theory) to an event, a biased belief may assign a 60\% chance to it (both are greater than 50\%). However, a posterior error is the consequence of a severe bias that switches the direction of the posterior probability, i.e., an updated belief that is in a wrong direction (e.g., a belief of less than 50\% in the earlier example).}

\item \textbf{Reasoning Error}: It might be that the subject's posterior is correctly in favor of state $X$, $Pr(X|x)>Pr(Y|x)$, but she mistakenly chooses state $Y$.
\end{enumerate}

In general, it is not possible to non-parametrically identify these two explanations by only observing the subject's choice. This implies that to overcome the identification challenges, one need extra variation in the data or additional information about subject's behavior. As stated before, my experimental design solves this identification problem by collecting data on both subjects' choices and their self-reported posteriors. So, I can distinguish between a posterior error and a reasoning error in the data.\footnote{The framework that is introduced here applies to both the individual condition and the social condition of my experiment. The only difference is that the signal (input) is a ball in the individual condition, while it is the guess of a neighbor (and any additional information coming along the neighbor's choice) in the social condition.}

\begin{figure}[htbp]
\centering
    \includegraphics[scale=0.6]{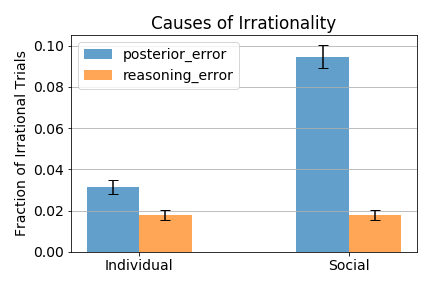}%
    \caption{The break down of individual irrationality and social irrationality into posterior error and reasoning error}
    \label{Ind-vs-Soc-breakdown-within_corrected}%
      
\end{figure}

Figure \ref{Ind-vs-Soc-breakdown-within_corrected} illustrates the break down of the observed individual irrationality and social irrationality into posterior error and reasoning error. The figure shows that the probability of a reasoning error is equal to 0.018 in both the individual condition and the social condition. However, there is a statistically significant difference between the magnitude of posterior error across the two conditions; it is 0.031 in the individual condition, but 0.095 in the social condition ($p\text{-}value<0.01$).

Figure \ref{Ind-vs-Soc-breakdown-within_corrected} provides an important insight about the implication of the violation of rational expectations in the social interaction. It suggests that the uncertainty about the neighbor's behavior makes the belief updating more difficult in the social condition than in the individual condition. That is, subjects make more errors when they update their beliefs (the first stage of the decision-making process) in the social condition than in the individual condition. However, this uncertainty does not influence the second stage of the decision-making process, i.e., once posterior beliefs are formed, subjects follow the same reasoning procedure, and thus the magnitude of the reasoning error remains unchanged across the individual condition and the social condition.

Distinguishing between posterior errors and reasoning errors identifies a critical touchpoint in the decision-making process and may provide solutions to nudge individuals towards making better decisions in social environments. It suggests that the violation of rational expectations may not necessarily result from the lack of Math/Stats knowledge, but may be more about how uncertain subjects are about their neighbor's behavior. In addition, failing to control for individual errors that are independent of the social environments (e.g., reasoning errors) may lead to unintended consequences. In appendix C, I estimate a unified model of individual behavior by borrowing techniques from the social learning literature \shortcite{Grether80,AndersonHolt97}. There, I show that not accounting for different sources of error (posterior error versus reasoning error) can bias the estimated structural parameter that intends to describe the individual behavior.\footnote{The predominant approach in the social learning literature is to derive predictions under the assumption that all other players obey a given model solution, for instance Perfect Bayesian Equilibrium or Quantal Response Equilibrium. But despite their undisputable usefulness, these solutions are often inaccurate descriptions of behavior and thus yield imperfect benchmarks \shortcite{Weizsacker10}.} 

\section{Non-parametric Estimation of the Decision-making Process}
\label{estimation}

In the previous section, I introduced a decision-making framework to explain how subjects behave in social interactions. In this section, I formally estimate a model that corresponds to this two-stage decision-making process. This model elaborates on three sources of irrationality under social interaction and sheds light on what variation in the data identifies which error. The estimation provides two non-parametric functions for each condition (individual/social): one function represents the belief updating process in stage 1, and the other function represents the choice rule in stage 2. For the first stage, I use the self-reported belief data, and for the second step, I use the combination of the choice and the belief data. The model is separately estimated for the individual condition and the social condition. 

 The identification of the sources of irrationality follows from the comparison of irrational choices between decision stages (stage 1 vs 2) and across conditions (individual vs social). The idea is that posterior errors, which are only present in stage 1, are often a combination of belief updating biases (type A) and violation of rational expectations (type C). In addition, we know from the earlier results, Figure \ref{Ind-vs-Soc-breakdown-within_corrected}, that the violation of rational expectations only exists in the social condition. So, the comparison of belief functions in stage 1 across the individual and the social condition separates type A error from type C error. However, reasoning errors (type B) are only present in stage 2. So, the choice function in stage 2 identifies type B errors. Table \ref{tab:identification} summarizes the identification insights from this section. Note that it was shown earlier that reasoning errors remain unchanged across the individual and social condition so that the comparison of stage 2 across conditions does not identify any new effect. If one believes there are other errors in the second stage which might be related to the social environment, for example due to the specifics of the context, then those errors can be also identified from this comparison. \\

\begin{table}[htbp]
\centering
\begin{tabular}{|c|c|c|}
\hline
                              & \textbf{Stage 1 (belief)} & \textbf{Stage 2 (choice)} \\ \hline
\textbf{Individual} & type A           & type B           \\ \hline
\textbf{Social}     & type A + type C  & type B           \\ \hline
\end{tabular}
\caption{The identification of three types of errors}
\label{tab:identification}
\end{table}

\subsection{Beliefs in Stage 1}
\begin{figure}[htbp]
\centering
    \includegraphics[scale=0.8]{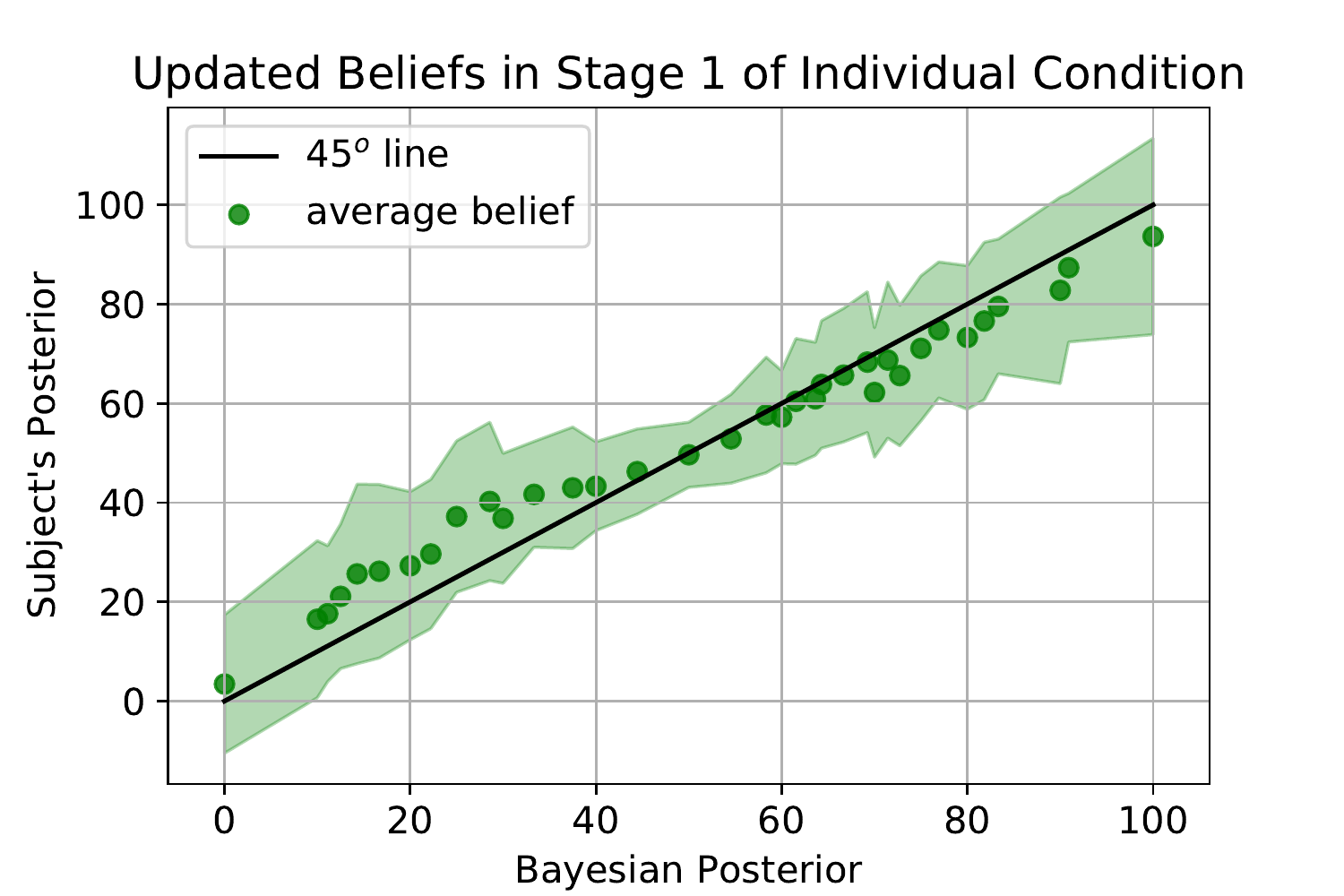}%
    \caption{The belief distribution across subjects in the individual condition}
    \label{fig:belief_indiv}%
\end{figure}

\noindent \textbf{\textit{Individual Condition.}} Denote subject $i$'s posterior belief about state $X$ after observing signal $s$ by $\mu_i(X|s;\theta)$, where $\theta = (\theta_X,\theta_Y)$ is the information structure and is common knowledge. I only assume $\mu_i(X|s;\theta)=1-\mu_i(Y|s;\theta)$ and do not impose any other assumption on $\mu_i$ so that the belief is individual-specific and as general as possible.\footnote{The well known model of \shortciteA{Grether80} is a special case of my model in which $\mu$ is the same across subjects (no heterogeneity) and it has a specific parametric form: $\mu = \frac{Pr(s|X;\theta)^c}{Pr(s|X;\theta)^c+Pr(s|Y;\theta)^c}$.} A unique aspect of the experiment is that it contains a wide variety of information structures so that there is enough variation in the probability space and I am able to plot the distribution of beliefs across subjects at different points. Figure \ref{fig:belief_indiv} presents the distribution of these beliefs ($\mu_i$) across subjects against the Bayesian posterior.\footnote{Note that the shaded area shows one standard deviation above and below the mean. In the data, no subject reports probabilities under 0 and over 100. However, when one plots the mean values with one standard deviation around it, the shaded area may contain probabilities out of $[0,100]$ range in corner points. This shall not mislead the reader.}

If subjects were Bayesian, one would expect all points to be exactly on the 45-degree line and the standard deviation would be zero. The first fact is that posterior beliefs are positively correlated with Bayesian beliefs. So, subjects generally understand the changes in likelihood. However, one can see over-appreciation of probabilities in the left hand side and under-appreciation of probabilities in the right hand side. The average self-reported posterior follows a trend which in essence is similar to the well-known probability weighting transformation in Cumulative Prospect Theory \shortcite{CPT}. This pattern is the result of what was called type A error in belief updating. Most importantly, there is substantial heterogeneity in beliefs so that the standard deviation is non-negligible and large specifically at boundaries (0 and 100). This is surprising because it implies some subjects make errors even in the extreme cases of certainty. So, to study the errors in the social interactions, one needs to carefully account for this type of heterogeneity.  \\

\begin{figure}[htbp]
\centering
    \includegraphics[scale=0.8]{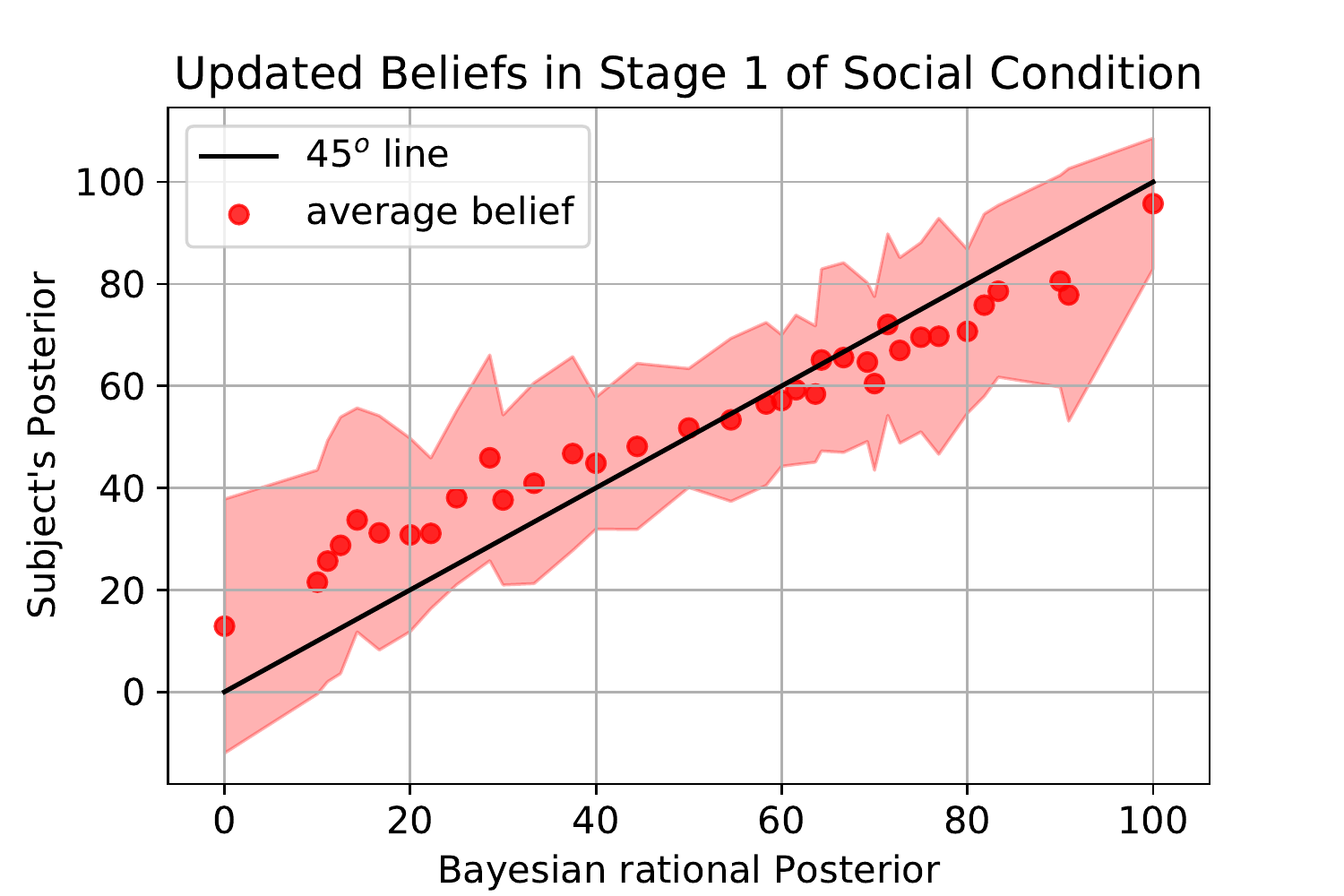}%
    \caption{The belief distribution across subjects in the social condition}
    \label{fig:belief_social}%
\end{figure}

\noindent \textbf{\textit{Social Condition.}} Denote the subject's posterior belief about state $X$ after observing a neighbor's guess of $g$ by $\gamma_i(X|g;\theta)$, where $\theta = (\theta_X,\theta_Y)$ is common knowledge. I impose only one assumption on this function: $\gamma_i(X|g;\theta)=1-\gamma_i(Y|g;\theta)$. Figure \ref{fig:belief_social} presents the distribution of these beliefs ($\gamma_i$) across subjects against the Bayesian rational posterior. The overall trend here is similar to that of individual condition. There is over-appreciation in the left side and under-appreciation in the right side. One difference between Figures \ref{fig:belief_indiv} and \ref{fig:belief_social} is that the variance is higher in the social condition than the individual condition. This is probably due to higher uncertainty included in the social condition. Recall that beliefs in the social condition are impacted by type A and type C errors. So, higher variance would be a natural implication in the social condition. 

\begin{figure}[htbp]
\centering
    \includegraphics[scale=0.8]{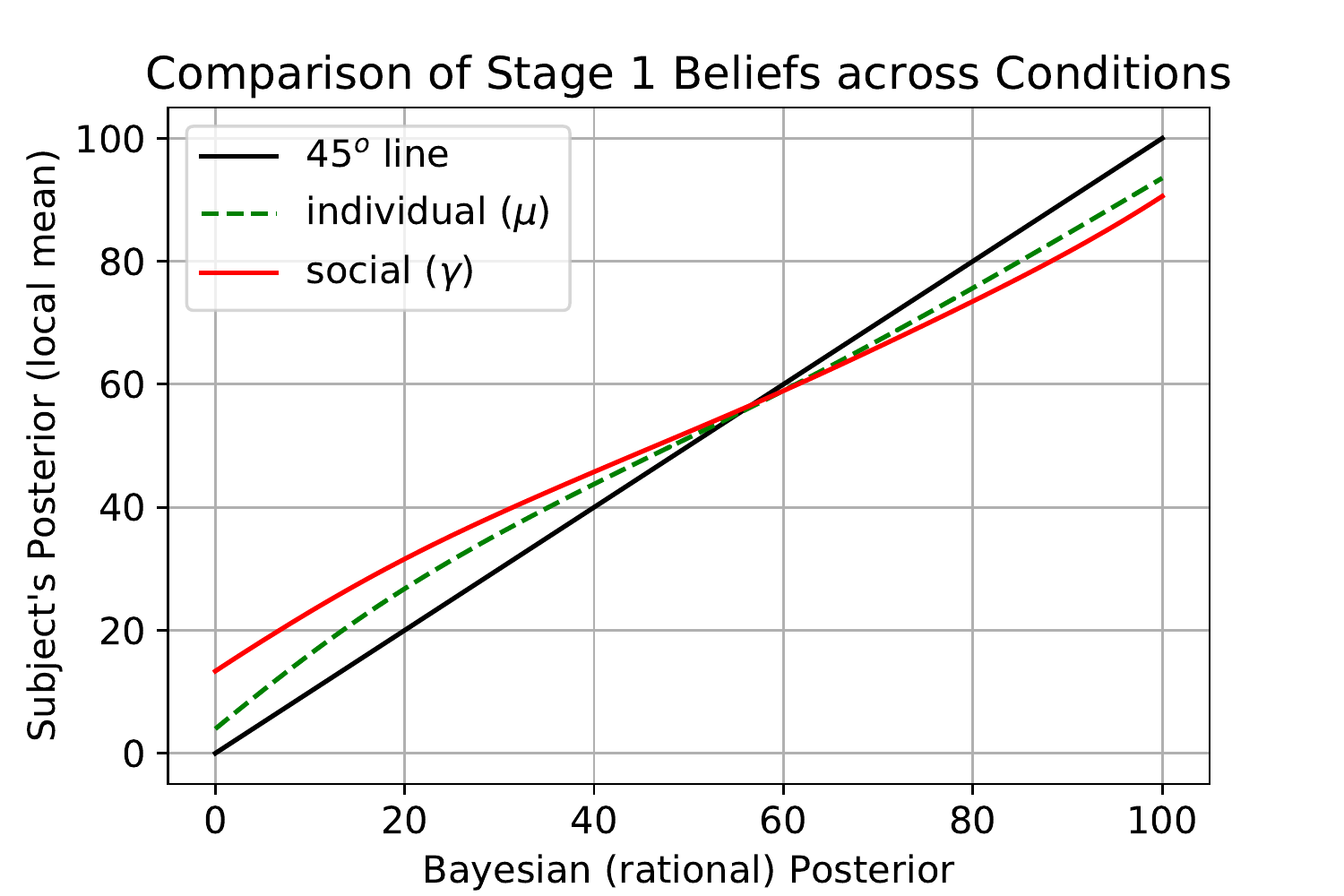}%
    \caption{The non-parametric estimation of mean posterior belief in the individual and social conditions}
    \label{fig:mu_gamma_mean}%
\end{figure}

To further separate type A and type C errors one needs to compare the average beliefs across conditions. One can see from Figures \ref{fig:belief_indiv} and \ref{fig:belief_social} that the average distance to 45-degree line seems to be higher in the social condition than in the individual condition. To further clarify on this comparison, I separately and non-parametrically estimate a belief function for each condition. The results of the Nadaraya-Watson kernel regression (bandwidth = 15) is presented in Figure \ref{fig:mu_gamma_mean}. As expected, the belief curve for the social condition exhibits a higher probability weighting bias compared to the individual condition. This effect is a result of what was earlier named as type C error, i.e., the violation of rational expectations.

\subsection{Choices in Stage 2}
\begin{figure}[htbp]
\centering
    \includegraphics[scale=0.8]{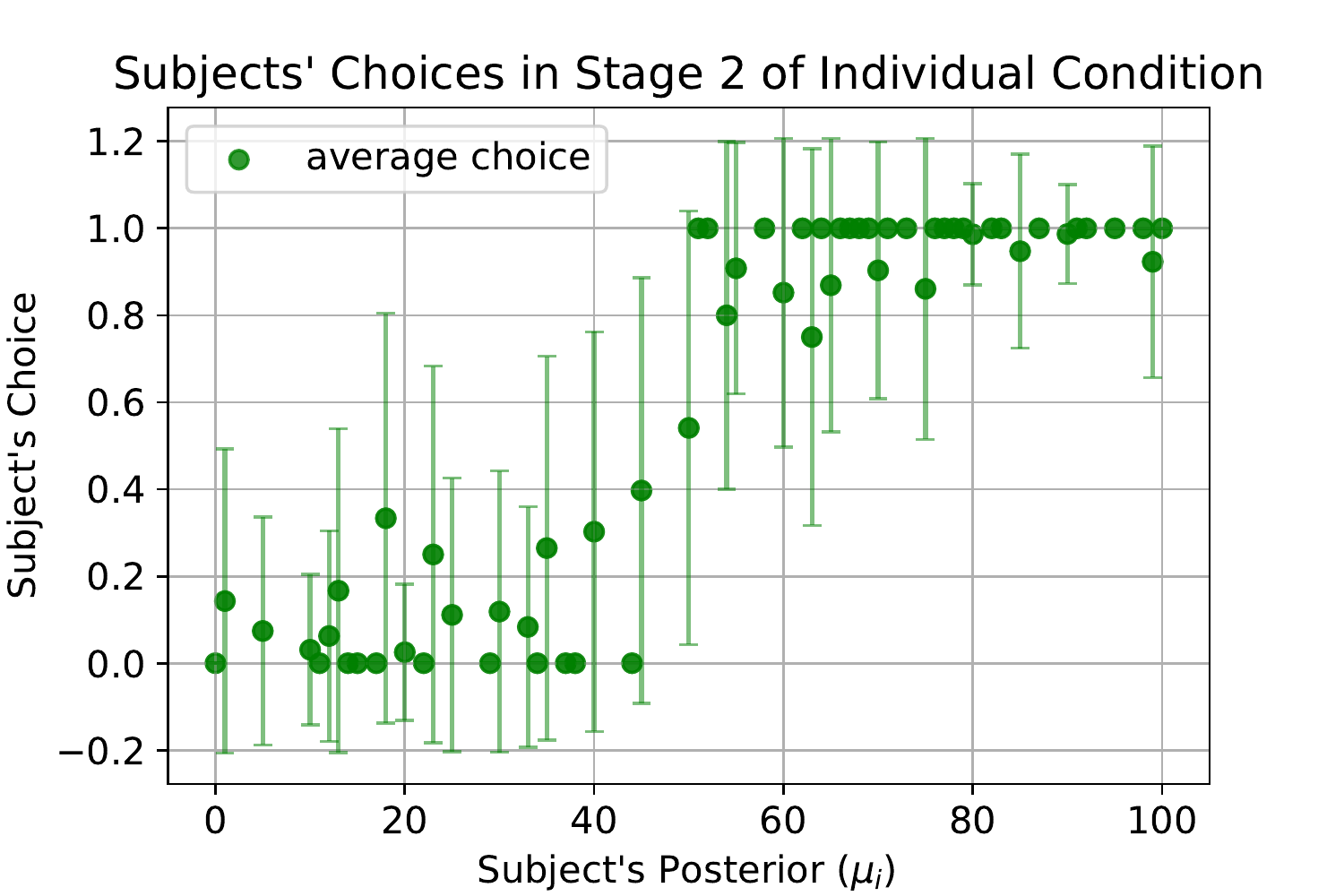}%
    \caption{The choice distribution across subjects in the individual condition}
    \label{fig:choice_indiv}%
\end{figure}

\noindent \textbf{\textit{Individual Condition.}} In the second step of decision-making, conditional on the posterior belief, $\mu_i$, the subject makes a binary decision. In the most general form, the subject's decision can be modeled as a mixed-strategy on $X$ and $Y$: she chooses state $X$ with probability $\alpha_i(\mu_i)$ and state $Y$ with probability $1-\alpha_i(\mu_i)$. Note that the standard Quantal Response Equilibrium \shortcite{McKelveyPalfrey95} imposes a parametric assumption on this so that $\alpha_i(\mu_i)$ is a logistic function. Also, Perfect Bayesian Equilibrium is a special case where $\mu_i$ is the Bayesian posterior and $\alpha_i$ is degenerate on the state with higher posterior probability. I combine the choice data from the second stage of the experiment with the belief data from the first stage and plot the distribution of $\alpha_i$ at each point in Figure \ref{fig:choice_indiv}.

If subjects were fully rational, all the points with $\mu_i<50$ would be on $Choice=0$ and all the points with $\mu_i>50$ would be on $Choice=1$, and the standard deviation would be zero at all points. The first general observation from Figure \ref{fig:choice_indiv} is that subjects respond to changes in posteriors, meaning that they are generally more likely to choose an option that has a higher probability. However, the positive standard errors demonstrate that, as discussed earlier, subjects may sometimes make reasoning errors, i.e., they may pick an option that they believe has a lower probability. This pattern can be consistent with a logistic function but one should take careful considerations regarding the substantial heterogeneity in choices. This analysis identifies the so called type B errors in subject's behavior.  

\begin{figure}[htbp]
\centering
    \includegraphics[scale=0.8]{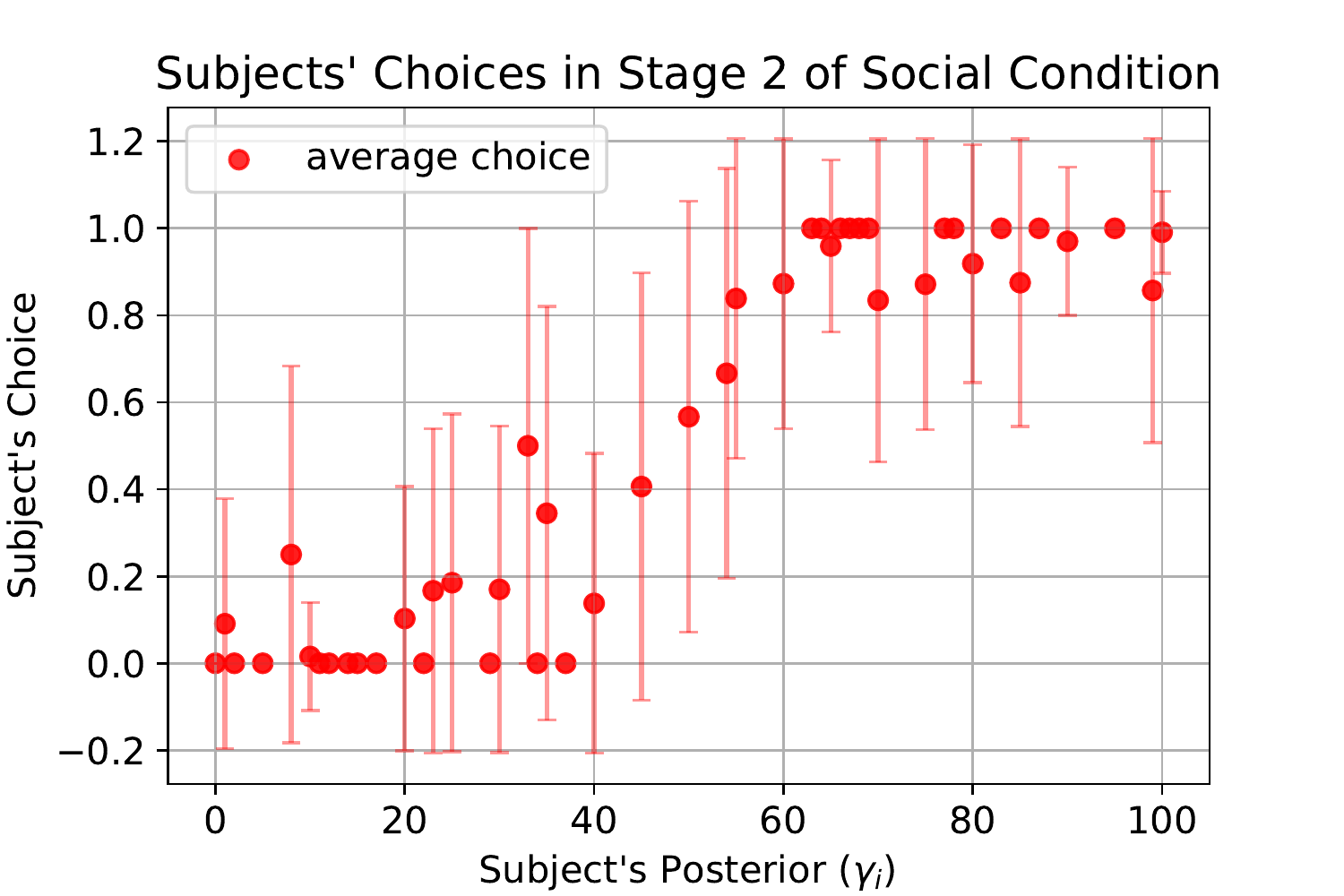}%
    \caption{The choice distribution across subjects in the social condition}
    \label{fig:choice_social}%
\end{figure}

\noindent \textbf{\textit{Social Condition.}} In the second stage of the social condition, the subject makes a binary decision based on the neighbor's choice. The subject's decision is a mixing probability on $X$ and $Y$: conditional on her posterior belief, $\gamma_i$, the subject chooses state $X$ with probability $\beta_i(\gamma_i)$ and state $Y$ with probability $1-\beta_i(\gamma_i)$. The distribution of $\beta_i$ is shown in Figure \ref{fig:choice_social}.

\begin{figure}[htbp]
\centering
    \includegraphics[scale=0.8]{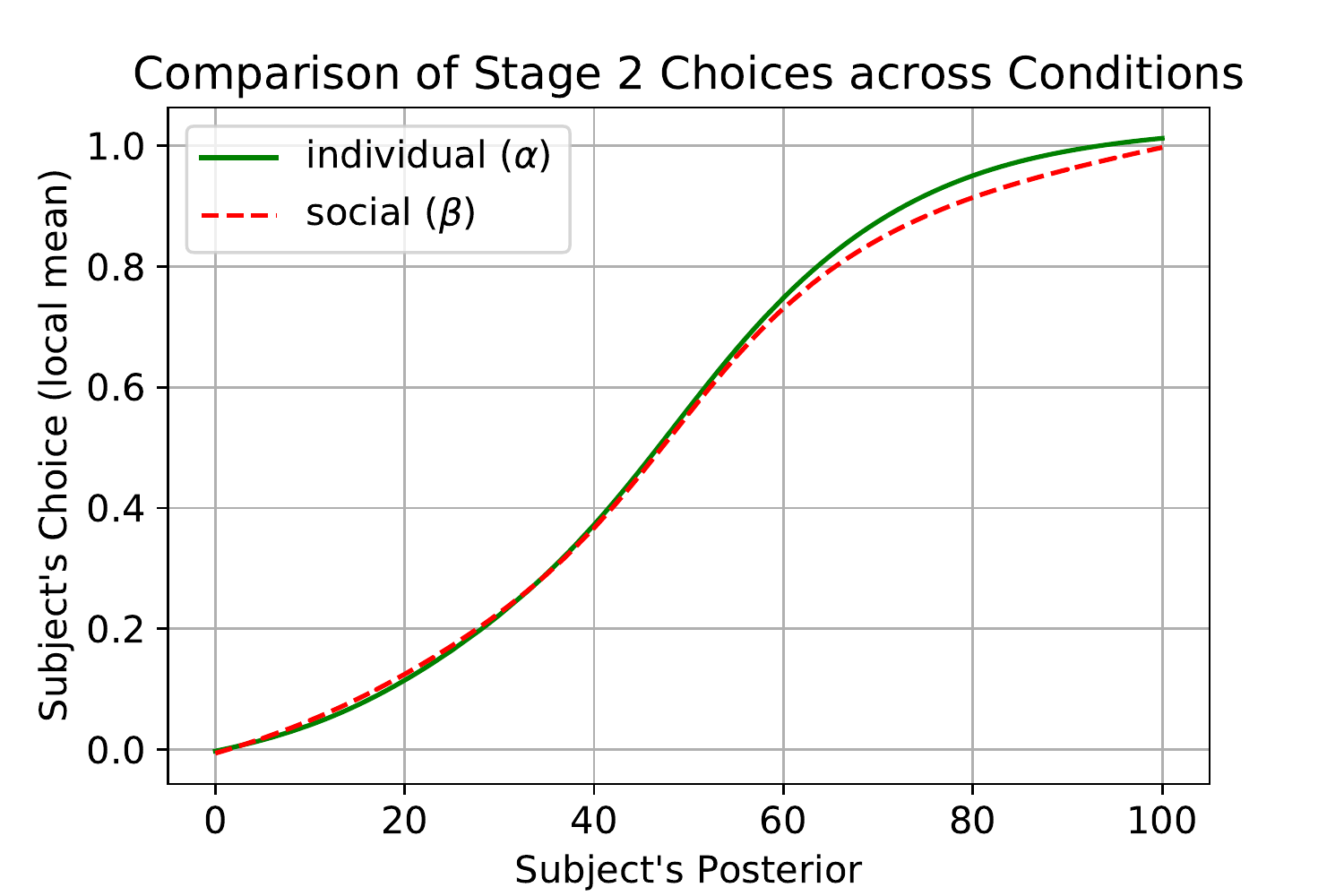}%
    \caption{The non-parametric estimation of choice probabilities in the individual and social conditions}
    \label{fig:alpha_beta_mean}%
\end{figure}

The general pattern in Figure \ref{fig:choice_social} is similar to that of the individual condition (Figure \ref{fig:choice_indiv}). However, to further check for the differences, I separately and non-parametrically estimate a choice function for each condition. The results of the Nadaraya-Watson kernel regression (bandwidth = 15) is presented in Figure \ref{fig:alpha_beta_mean}. As expected, the choice curve for the social condition exhibits a very similar pattern to that of the individual condition.\footnote{Note that, as I mentioned earlier, the experimental design is such that it can also identify errors related to SI that happen in the second stage of decision-making and are separate from the other three types of error. So, one might argue that the small difference between choice probabilities in the right hand side of Figure \ref{fig:alpha_beta_mean} is due to such errors. However, I believe this small difference is not due to a systematic error. Recall that in the previous section, I showed that the average reasoning error remains unchanged across conditions. So, I think the difference here might be just because of luck or may be related to the noise in the data (here, the analysis is at a more granular level than the previous section).}

\section{Conclusions and Implications}
\label{conclusion}

Many economic decisions, from the most mundane ones, like choosing a diner for lunch, to the most important ones, like adoption of a new technology by a firm, require making inferences about an underlying state of the world. In these situations, social interaction is an important source of information. Economic agents often interact with each other via observation of choices. Such social interactions affect people's beliefs and can help them to make informed decisions. 

The conventional assumption in the economic models is that decision makers are Bayesian and they have rational expectations about each other. However, individuals often deviate from these assumptions. Errors in social interactions may have severe consequences, for example, they may lead to a socially inefficient equilibrium when information is transmitted by observation. So, it is important to study and identify the sources of such errors.
  
In this study, I conduct a series of laboratory experiments to uncover errors in learning from others' choices. I use a relatively simple and novel experimental setting to disentangle between individual errors that are independent of the social environment, and the errors that are caused by the violation of rational expectations. In a within-subject design, I compare subjects' choices across an isolated condition and a social condition, and show that subjects make more errors in the presence than in the absence of social interaction, even when they receive informationally equivalent signals across the two conditions. That is, they neglect the provided information more when they interact with others than when they do not. 

To uncover the mechanism behind the additional neglect in the social condition, I design a series of treatment variations by exogenously manipulating the subject's knowledge about her neighbor. I find that the unexpected irrationality in the social condition is mainly driven by the uncertainty about other people's behavior: subject's behave as if they lack knowledge of how others make decision based on their private signals. The implication of this result is that social interactions might not be as effective as one expects in theory. So, one should take careful considerations in examining the effects of social interactions using field data.

In econometric models of field data, researchers often impose the assumption of Bayesian rationality. While one should understand the need for these assumptions, to overcome identification challenges, it has been unclear whether and how one can identify different types of errors in decision making under observational learning. By introducing a unique experimental design, this paper sheds lights on three categories of error that often emerge in social interactions: errors associated with belief updating, errors due to incorrect reasoning, and errors related to the violation of rational expectations. I show that measuring probabilistic beliefs is important because it provides extra data which can be utilized to identify new aspects of individual behavior. However, to non-parametrically estimate the complete decision-making process, one needs extra variations in the data. While the estimation is specific to the context of this study, similar analysis can be used to separate various errors in decision-making using field data. Identification of non-equilibrium beliefs is an important topic that has recently attracted substantial attention \shortcite{VictorArvind20,VictorErhao21}. This study extends this literature to situations where there is no strategic incentives between players and identifies the contribution of different errors to subjects' behavior. The results of this study can provide guidance to researchers and managers on how to benefit from measuring beliefs and incorporate them into their analysis.

In many circumstances, marketers may observe inefficiencies in the market behavior. For example, in the kidney market, kidney utilization may remain low despite the continual shortage in kidney supply. The results of this study highlight the role of decision-making errors in such situations. Understanding the sources of error can guide policy-makers to design policies that nudge individuals towards socially optimal outcomes. 



\newpage
\bibliographystyle{apacite}
\bibliography{mylib}

\newpage 

\section*{Appendix A1: Experiment Instructions}
Welcome and thank you for participating in this experiment. Please read this instruction carefully. All participants in this experiment are recruited in the same way as you and read the same instructions as you do. 
It is important that you do not discuss the details of this experiment with anyone else after the experiment. 

You will receive \$6 for participating in this experiment. During the course of the experiment you can earn more. Your earning will depend on your performance during the experiment. All your decisions and your earning will be treated confidentially.

This session is part of an experiment about how humans make decision under uncertainty. The experiment consists of multiple parts. You need to read the following instruction to understand what will happen in each part of the experiment.\\
\\
\underline{\textbf{FIRST PART:}}\\
In the first part of the experiment, you'll face 21 rounds of decision-making. Each round proceeds as follows:
\\
Two boxes are shown to you. Each box contains 10 balls. The combination of BLACK and WHITE balls can be different in each round. You can see an example below.
\begin{figure}[htbp]
\centering
        \includegraphics[scale=0.8]{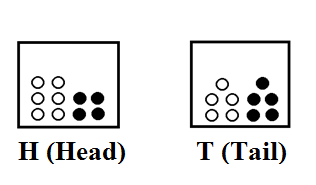}
\end{figure}

Then, a fair coin is flipped. If the coin is \textbf{Tail}, one ball is randomly drawn from \textbf{box T (Tail)}. If the coin is \textbf{Head}, one ball is randomly drawn from \textbf{box H (Head)}. You do not observe the coin, but you see the ball. For example if the ball is black, you see the following:
\\
\\
\begin{figure}[htbp]
\centering
        \includegraphics[scale=0.8]{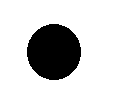}
\end{figure}

You then guess which box the ball has been drawn from. Specifically, you will answer the following question: 
\begin{figure}[htbp]
\centering
        \includegraphics[scale=0.9]{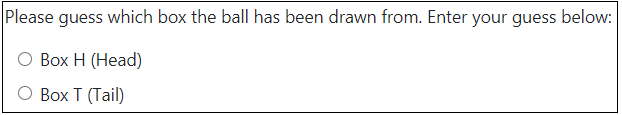}
\end{figure}

Remember that your answers during the experiment determine your payment at the end of the experiment. So, if you guess the correct box, you will likely receive more money at the end of the experiment. At the end of the experiment, two of your guesses will be selected at random and you will receive \$12 for each correct guess.

To summarize part 1 of the experiment, you will face 21 rounds. In each round, you experience the following steps:

\begin{enumerate}
    \item  A fair coin is flipped. If the coin is Tail, one ball is randomly drawn from box T (Tail). If the coin is Head, one ball is randomly drawn from box H (Head).
    
    \item You see the ball.
    
    \item You guess which box the ball has been drawn from.
\end{enumerate}
Note that you should consider each round as an independent experiment. There is no relation between different rounds.
\\
\\
\underline{\textbf{SECOND PART: (Treatment \textit{Base})}}\\
In the second part of the experiment, you again go through 21 rounds. Each round proceeds as follows: \\
You are randomly connected to another participant, who is called your "NEIGHBOR". Your neighbor is an individual who has participated in this experiment before. He/She is not present in the lab right now. You then receive information about one of the rounds that your neighbor has gone through. The setting of your neighbor's experiment was the same as what you saw in the previous part. In each round, the neighbor was observing one ball, randomly drawn from one of the two available boxes, and then he/she was submitting a guess.

Your task is as follows: you first see the two boxes that has been shown to your neighbor. You then observe your neighbor's guess. Finally, you guess what the correct box is in your neighbor's experiment. Remember that your neighbor's guesses were for money. It means that a correct guess by your neighbor was increasing his/her chances of getting money. 

To summarize the second part of the experiment, you will face 21 rounds. In each round, you are randomly connected to another participant, who is called your "NEIGHBOR". Then, you proceed as follows:
\begin{enumerate}
    \item You see two boxes which were shown to your neighbor.
    \item You see your neighbor's guess (your neighbor's guess was based on a ball randomly drawn from one of the boxes).
    \item You guess which box is the correct choice in your neighbor's experiment.
\end{enumerate}
Please keep in mind that you perform this task in multiple rounds and you might be connected to a different person in each round. So, you are not necessarily interacting with the same person in all rounds.
\\
\\
\underline{\textbf{SECOND PART: (Treatment \textit{Demographics})}}\\
In the second part of the experiment, you again go through 21 rounds. Each round proceeds as follows:
\\
You are randomly connected to another participant, who is called your "NEIGHBOR". Your neighbor is an individual who has participated in this experiment before. He/She is not present in the lab right now. You then receive information about one of the rounds that your neighbor has gone through. The setting of your neighbor's experiment was the same as what you saw in the previous part. In each round, the neighbor was observing one ball, randomly drawn from one of the two available boxes, and then he/she was submitting a guess.

Your task is as follows: you first see the two boxes that has been shown to your neighbor. You then observe your neighbor's guess (you will also see some demographic information about your neighbor such as: gender, age, years of education, and whether he/she has taken any Probability/Statistics course). Finally, you guess what the correct box is in your neighbor's experiment. Remember that your neighbor's guesses were for money. It means that a correct guess by your neighbor was increasing his/her chances of getting money. 

To summarize the second part of the experiment, you will face 21 rounds. In each round, you are randomly connected to another participant, who is called your "NEIGHBOR". Then, you proceed as follows:
\begin{enumerate}
    \item You see two boxes which were shown to your neighbor.
    \item You see your neighbor's guess (your neighbor's guess was based on a ball randomly drawn from one of the boxes). You also observe some demographic information about your neighbor.
    \item You guess which box is the correct choice in your neighbor's experiment.
\end{enumerate}
Please keep in mind that you perform this task in multiple rounds and you might be connected to a different person in each round. So, you are not necessarily interacting with the same person in all rounds.
\\
\\
\underline{\textbf{SECOND PART: (Treatment \textit{Bot})}}\\
In the second part of the experiment, you again go through 21 rounds. Each round proceeds as follows:\\
You are connected to a bot, which is called your "NEIGHBOR". The setting of the experiment is similar to what you saw in the previous part. Two boxes are shown to you and the bot. Then, a ball is randomly drawn from one of the boxes. We show the ball to the bot (not you) and let the bot guess which box the ball is drawn from. 

The bot is programmed such that when it observes a white ball, it picks the box with more white balls, and when it observes a black ball, it picks the box with more black balls. You then see the bot's guess. Finally, you guess which box the ball has been drawn from.

To summarize the second part of the experiment, you will face 21 rounds. In each round, you are connected to a bot that is called your "NEIGHBOR". Then, you proceed as follows:
 \begin{enumerate}
     \item You and your neighbor (BOT) see two boxes.
     \item You see your neighbor's guess (your neighbor's guess is based on a ball randomly drawn from one of the boxes).
     \item You guess which box is the correct choice.
 \end{enumerate}
NOTE: The bot is programmed such that it picks the box with more black balls when it observes a black ball, and it picks the box with more white balls when it observes a white ball.
\\
\\
\underline{\textbf{SECOND PART: (Treatment \textit{Ball})}}\\
In the second part of the experiment, you again go through 21 rounds. Each round proceeds as follows:
\\
You are randomly connected to another participant, who is called your "NEIGHBOR". Your neighbor is an individual who has participated in this experiment before. He/She is not present in the lab right now. You then receive information about one of the rounds that your neighbor has gone through. The setting of your neighbor's experiment was the same as what you saw in the previous part. In each round, the neighbor was observing one ball, randomly drawn from one of the two available boxes, and then he/she was submitting a guess.

Your task is as follows: you first see the two boxes that has been shown to your neighbor. You then observe your neighbor's guess and the ball that has been shown to him/her. Finally, you guess what the correct box is in your neighbor's experiment. Remember that your neighbor's guesses were for money. It means that a correct guess by your neighbor was increasing his/her chances of getting money. 

To summarize the second part of the experiment, you will face 21 rounds. In each round, you are randomly connected to another participant, who is called your "NEIGHBOR". Then, you proceed as follows:
\begin{enumerate}
    \item You see two boxes which were shown to your neighbor.
    \item You see your neighbor's guess as well as the ball that had been shown to your neighbor.
    \item You guess which box is the correct choice in your neighbor's experiment.
\end{enumerate}
Please keep in mind that you perform this task in multiple rounds and you might be connected to a different person in each round. So, you are not necessarily interacting with the same person in all rounds.
\\
\\
\underline{\textbf{PAYMENT:}}\\            
You will receive \$6 for showing up. In addition to that, two of your guesses during the experiment will be randomly selected and you get an extra \$12 for each correct guess.
\\
Note: Your payment will be determined at the end of the experiment session after you finish all parts of the experiment. You will not know how much you earn in each part during the experiment.

\newpage

\section*{Appendix A2: Experiment Interface (oTree)}

\begin{figure}[htbp]
\centering
        \includegraphics[scale=0.63]{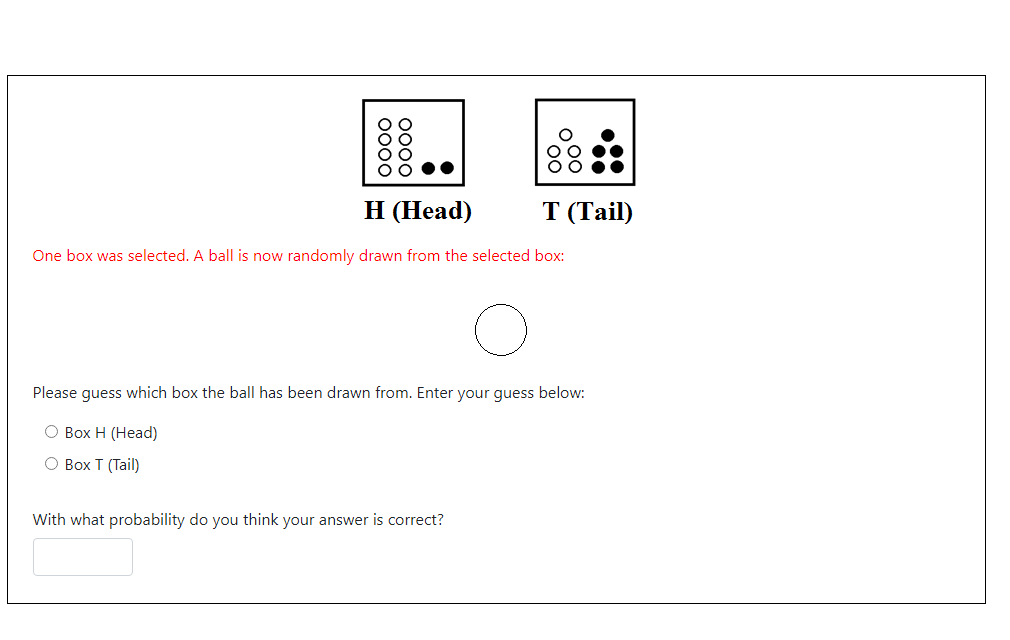}
      \caption{Individual Condition}
\end{figure}
\newpage
\begin{figure}[htbp]
\centering
        \includegraphics[scale=0.63]{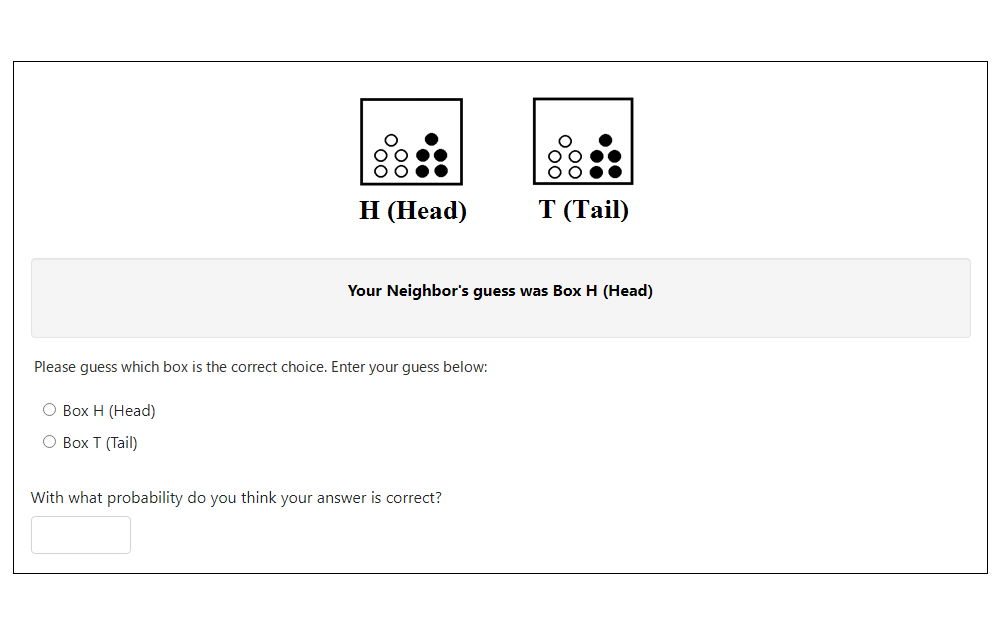}
      \caption{Social Condition (Base)}
\end{figure}
\newpage
\begin{figure}[htbp]
\centering
        \includegraphics[scale=0.71]{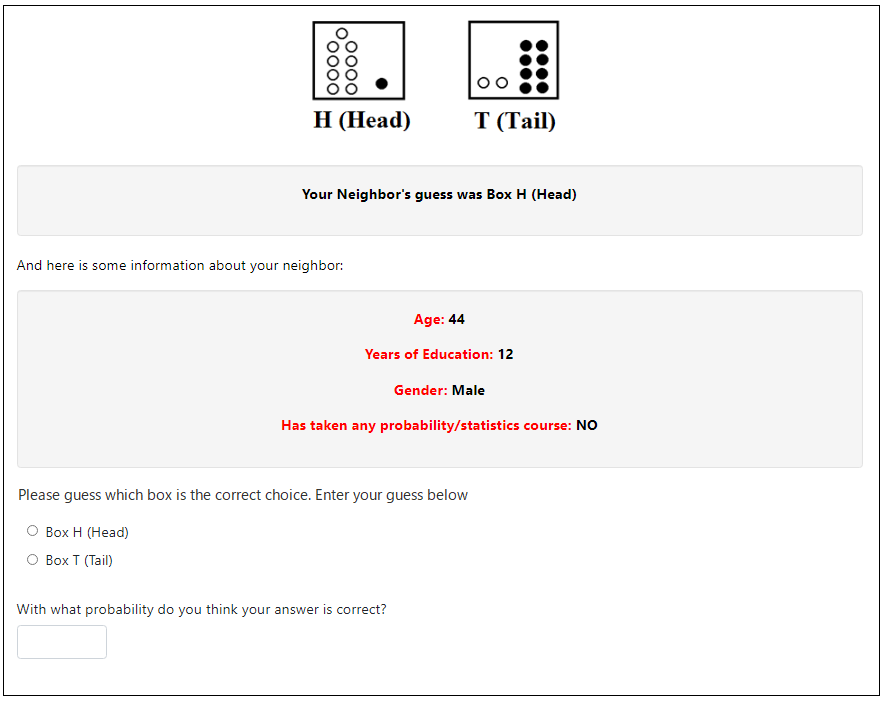}
      \caption{Social Condition (Demographics)}
\end{figure}
\newpage
\begin{figure}[htbp]
\centering
        \includegraphics[scale=0.70]{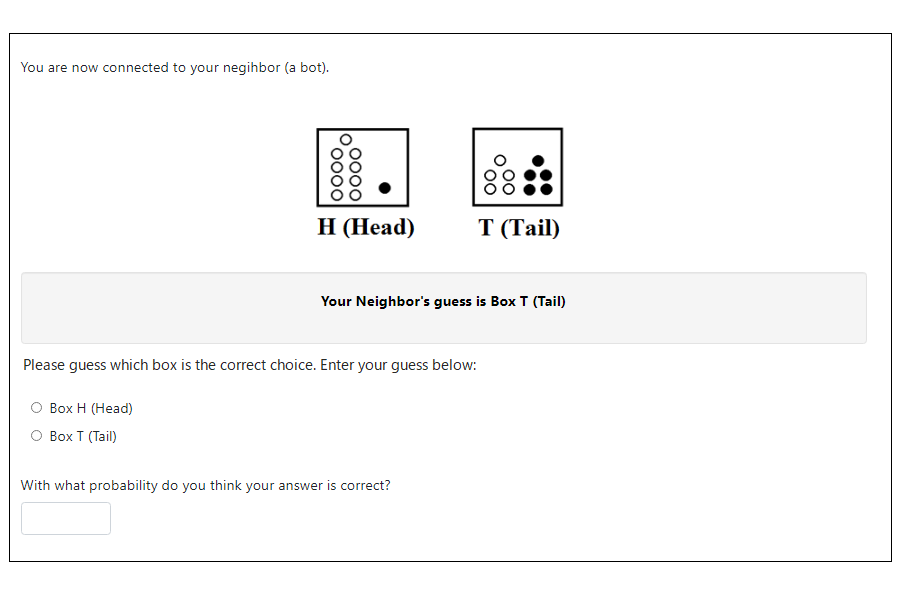}
      \caption{Social Condition (Bot)}
\end{figure}
\newpage
\begin{figure}[htbp]
\centering
        \includegraphics[scale=0.72]{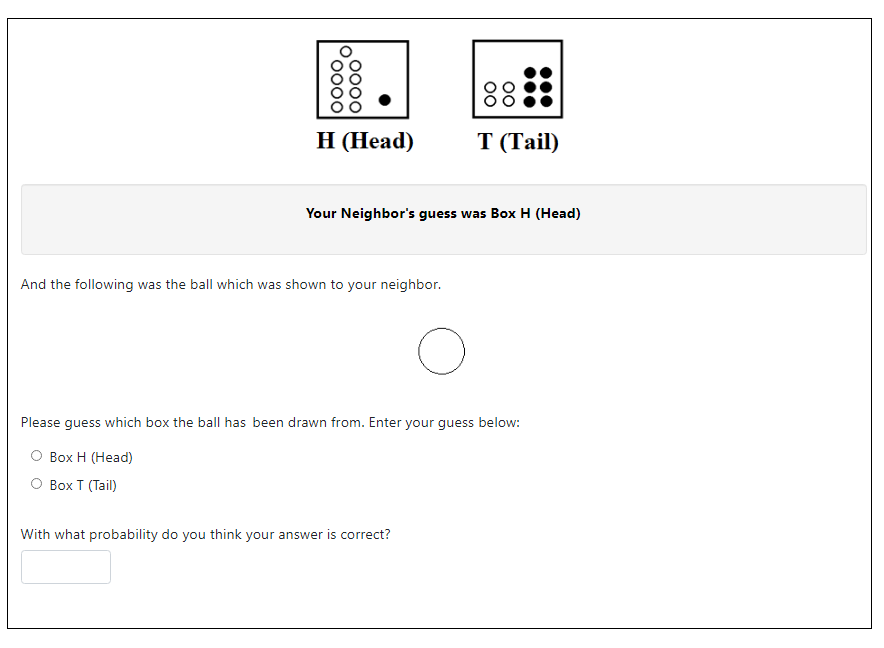}
      \caption{Social Condition (Ball)}
\end{figure}

\newpage

\section*{Appendix B: Between-Subject Analysis}
\begin{figure}[htbp]
\centering
        \includegraphics[scale=0.6]{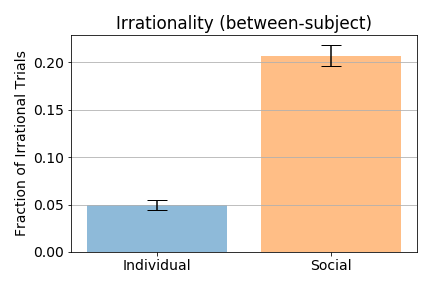}
      \caption{Individual Irrationality and Social Irrationality (between-subject analysis)}
      \label{Ind-vs-Soc-total-between-nolearning_corrected}
\end{figure}

In the previous sections, the analysis was done on a pooled data, meaning that the data for each subject consists of decisions in both the individual and the social conditions. This means the earlier results were based on a "within-subject" analysis. In this section, I do a robustness check and examine whether the results hold if the analysis is done "between-subject". To achieve this, I need to compare the individual irrationality of subjects who see the individual condition first, with the social irrationality of subjects who see the social condition first. The results of this between-subject analysis is shown in Figure \ref{Ind-vs-Soc-total-between-nolearning_corrected}. This figure verifies that the main finding holds even if the analysis is done \textit{between-subject}.

\begin{figure}[htbp]
\centering
        \includegraphics[scale=0.6]{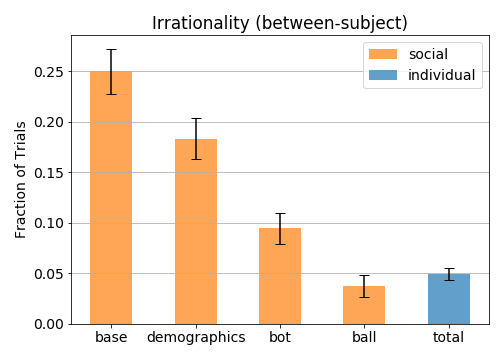}
      \caption{Social irrationality in different treatments (between-subject analysis)}
      \label{Ind-vs-Soc-nolearning_treatments_corrected}
\end{figure}

I can also redo the underlying analysis of Figure \ref{Ind-vs-Soc-treatments_corrected} at a between-subject level. The results, which are shown in Figure \ref{Ind-vs-Soc-nolearning_treatments_corrected}, are qualitatively similar to those presented in the body of the article.

\section*{Appendix C: A Unified Model of Individual Behavior}
\label{model}
In this section, I present and estimate a behavioral model that corresponds to the two-stage decision-making process introduced in the body of the article. The model combines two frameworks from the existing literature: for the first stage of the decision-making process, the model adapts a standard framework that was introduced in \shortciteA{Grether80}. For the second stage of the decision-making process, the model uses logistic response functions to determine choice probabilities \shortcite{McKelveyPalfrey95,AndersonHolt97}. The model parameters are estimated by a two-step method. In the fist step, I use the data from the individual condition to estimate the individual errors that are independent of the social condition. These estimates are then used in a second step to estimate the subject's belief about the neighbor's behavior in the social condition. This two step method uncovers the impact of the uncertainty about the neighbor's behavior on the subject's error rate in the social condition.

\subsection*{The Behavioral Model}
My goal in here is to build a comprehensive model that incorporates the two stages of the decision-making process shown in Figure \ref{decision process}. For the first stage, I use the traditional framework of \shortciteA{Grether80}. In this framework, when an individual obtains signal $s$, she updates her belief as follows:
\begin{align}
\pi(X|s)=\frac{p(s|X)^c \ p(X)}{p(s|X)^c \ p(X)+p(s|Y)^c \ p(Y)}    
\end{align}
where $s \in \{x,y\}$, $p(.)$ is the true probability, and $c$ is a measure of bias in the posterior. Bayes rule is a special case of this equation with $c=1$. In my experiment, priors are $p(X)=p(Y)=0.5$. So, the above equation can be simplified as follows,
\begin{align}
\label{c_eqn}
 \pi(X|s)=\frac{p(s|X)^c}{p(s|X)^c+p(s|Y)^c}
\end{align}
Here, $c<1$ is associated with underinference, i.e., as if the signals provide less information about the state than they actually do. Accordingly, the more the posterior errors, the lower the $c$. Alternatively, $c>1$ corresponds to overinference, signals provide more information than they actually do.

For the second stage of the decision-making process, I use a logit function \shortcite{AndersonHolt97}. Here, conditional on the posterior belief, $\pi(.|s)$, an individual makes a binary choice, $D$, with the following probability,
\begin{align}
\label{error_eqn}
Pr(D=X|s)=\frac{1}{1+e^{-\beta\big(\pi[X|s]-\pi[Y|s]\big) U}}=\frac{1}{1+e^{-\beta\big(2\pi(X|s)-1\big)U}}
\end{align}
where $\beta$ is a measure of response precision and $U$ is the bonus of a correct guess (\$12). In this framework, the reasoning error is inversely related to $\beta$: the reasoning error diminishes when $\beta \rightarrow \infty$, and it increases as $\beta\rightarrow 0$. Note that there is a notable difference between this model and that of \shortciteA{AndersonHolt97}. In \shortciteA{AndersonHolt97}, the subjects are assumed to update their beliefs using the Bayes rule, but here the posterior belief, $\pi(.|s)$, is not necessarily Bayesian (e.g., it can be biased).

\begin{figure}[htbp]
\begin{center}
\begin{tikzpicture}[xscale=1]
\draw[->][draw=black, very thick]  (-1,0)  node [left] {Input (signal)} -- (0.5,0);;
\draw[-][draw=black, very thick]  (0,0.75) -- (9,0.75);
\draw[-][draw=black, very thick]  (0,-0.75) -- (9,-0.75);
\draw[->][draw=black, very thick]  (3.5,0) -- (5.5,0);

\draw[->][draw=black, very thick]  (8.5,0) -- (10,0) node [right] {Output (choice)};
\draw [draw=black, very thick](0,-0.75)  -- (0, 0.75);
\draw [draw=black, very thick](9,-0.75)  -- (9, 0.75);

\node[align=left, color=purple] at (2,0)
{Update Belief};
\node[align=left, below, color = purple] at (2,-1)
{$\pi(X|s)=\frac{p(s|X)^c}{p(s|X)^c+p(s|Y)^c} $};
\node[align=center, above] at (4.5,1)
{Decision-Making Process};
\node[align=right, color=blue] at (7,0)
{Binary Decision};
\node[align=left, below, color = blue] at (8,-1)
{$Pr(D=X|s)=\frac{1}{1+exp\big[-\beta\big(2\pi(X|s)-1\big)U\big]}$};
\end{tikzpicture}
\end{center}
\caption{The behavioral model}
\label{processwitheqn}
\end{figure}
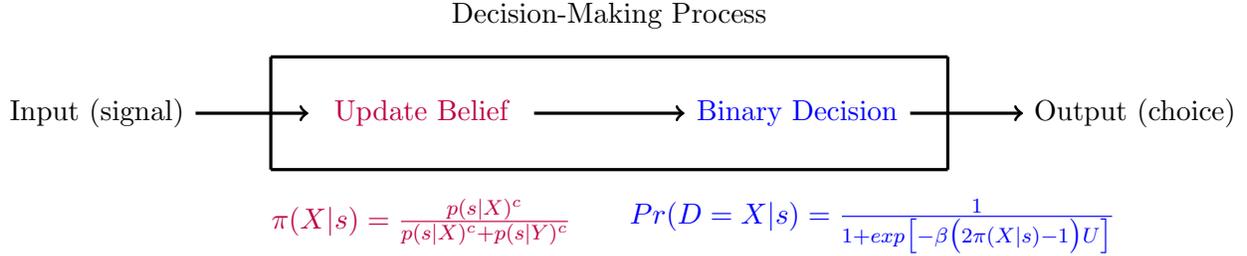

Figure \ref{processwitheqn} summarizes the behavioral model. In the first stage, the subject updates her belief according to equation \eqref{c_eqn}. In the second stage, conditional on the updated belief, the subject makes a binary choice according to equation \eqref{error_eqn}.

\subsection*{Estimation: First Step}
The first step of the estimation employs data from the individual condition of my experiment. In the individual condition, $s$ is the color of a ball randomly drawn from the realized state. So, the objective probabilities $p(s|X)$ and $p(s|Y)$ can be easily computed from the content of each box. In addition, the subject's choice, $D$, and her posterior, $\pi(X|s)$, are directly observed in the data. This implies that parameters $c$ in equation \eqref{c_eqn} and $\beta$ in equation \eqref{error_eqn} can be directly estimated from the data in the individual condition. For the parameter $c$, I first rearrange equation \eqref{c_eqn} as a linear regression, $\text{ln}\Big(\frac{\pi(X|s)}{\pi(Y|s)}\Big)=   c \ \text{ln}\Big(\frac{p(s|X)}{p(s|Y)}\Big)$, and then use OLS for the estimation.

The estimation results are reported in Table \ref{Regress_results}. Column (1) shows that the estimated value for parameter $c$ is significantly lower than 1, the 95\% confidence interval for $c$ is (0.872,0.903), meaning that in the individual condition, subjects infer less from the signal than they should. This evidence is consistent with the fact that subjects make posterior errors in the first stage of the decision-making process. Column (2) presents the estimated value for $\beta$. The small value of this parameter speaks to the fact that subjects make reasoning errors in the second stage of the decision-making process. 
\\
\begin{table}[ht] \centering 
  \caption{The First Step Estimation Results Using the Pooled Data of the \underline{Individual} condition} 
  \label{Regress_results} 
\begin{tabular}{@{\extracolsep{5pt}}lccc} 
\\[-1.8ex]\hline 
\hline \\[-1.8ex] 
\\[-1.8ex] & OLS & Logit:True Posterior & Logit:Bayes Posterior\\
\\[-1.8ex] & (1) & (2) & (3)\\
\hline \\[-1.8ex] 
 c & 0.888$^{***}$ & & \\ 
  & (0.008) &  &\\ 
  & & &\\ 
 $\beta$ &  & 0.472$^{***}$ & 0.373$^{***}$\\ 
  &  & (0.018) & (0.012) \\ 
\hline \\[-1.8ex] 
Observations & 3171 & 3171 & 3171\\ 
R$^{2}$ & 0.778 &  \\ 
Pseudo R$^{2}$ &  & 0.519 & 0.647\\ 
\hline 
\hline \\[-1.8ex] 
  & \multicolumn{2}{r}{$^{***}p<0.01$} \\ 
\end{tabular} 
\end{table} 
%
%

Column (3) in Table \ref{Regress} presents the estimated value for $\beta$ conditional on the assumption that posteriors in equation \eqref{error_eqn} are Bayesian. This is the conventional assumption in the social learning experiments \shortcite{AndersonHolt97,KublerWei04}. Here, the estimate for $\beta$ is 0.373, significantly lower than the estimated value in the second column. This comparison suggests that the posterior error is an important factor in the individual decision-making process and not accounting for posterior errors, biased beliefs in the logit function, can lead to biased estimates for $\beta$ (overestimation of the reasoning error).

\subsection*{Estimation: Second Step}
The goal in the second step of the estimation is to use the estimates from the first step, $\hat{c}=0.888$ and $\hat{\beta}=0.472$, to estimate the subject's belief about the neighbor's error rate in the social condition. Here, the estimation only employs data from the social condition of my experiment. I exclude the data for the \textit{ball} treatment in this section. As noted earlier, the subject observes both the neighbor's guess and the ball that the neighbor has observed. So, the subject's belief about the neighbor's error rate is irrelevant in that treatment.

Note that the decision-making process in the social condition still follows the framework in Figure \ref{processwitheqn}, with one exception: here, the subject observes the guess of a neighbor as signal. Denote the neighbor's guess by '$s$' $\in \{$'$x$','$y$'$\}$ and the neighbor's private signal by $s \in \{x,y\}$. The subject's posterior after observing the neighbor's guess can be derived as follows,
\begin{align}
\label{c_eqn_social}
 \pi(X|\text{'}s\text{'})=\frac{p(\text{'}s\text{'}|X)^c}{p(\text{'}s\text{'}|X)^c+p(\text{'}s\text{'}|Y)^c} = \frac{p(\text{'}s\text{'}|X)^{0.888}}{p(\text{'}s\text{'}|X)^{0.888}+p(\text{'}s\text{'}|Y)^{0.888}}
\end{align}

The parameter $c$ in this equation is a structural parameter and is set to 0.888, the estimate from the first step estimation. However, $p(\text{'}s\text{'}|X)$ and $p(\text{'}s\text{'}|Y)$ in equation \eqref{c_eqn_social} cannot be directly computed by observing the content of boxes. These probabilities depend on the subject's belief about the behavior of the neighbor. Let's denote the subject's belief about her neighbor's error rates by a pair $(\tilde{\beta},\tilde{c})$. Then, it is straightforward to derive the followings: 
\begin{align}
\label{social_specific}
    p(\text{'}x\text{'}|X)=Pr(D_n=X|x) \ p(x|X)+Pr(D_n=X|y) \ p(y|X)
\\
\label{social_specific2}
    p(\text{'}y\text{'}|X)=[1-Pr(D_n=X|x)] \ p(x|X)+[1-Pr(D_n=X|y)] \ p(y|X)
\\
\label{social_specific3}
    p(\text{'}x\text{'}|Y)=Pr(D_n=X|x) \ p(x|Y)+Pr(D_n=X|y) \ p(y|Y)
\\
\label{social_specific4}
    p(\text{'}y\text{'}|Y)=[1-Pr(D_n=X|x)] \ p(x|Y)+[1-Pr(D_n=X|y)] \ p(y|Y)
\end{align}
where $Pr(D_n=X|s)$, $s\in\{x,y\}$ is the subject's belief about the neighbor's choice probability:
\begin{align}
\label{belief about neighbor}
    Pr(D_n=X|s)=\frac{1}{1+e^{-\tilde{\beta}\big(2\tilde{\pi}(X|s)-1\big)U}}
\end{align}
and $\tilde{\pi}(X|s)=\frac{p(s|X)^{\tilde{c}}}{p(s|X)^{\tilde{c}}+p(s|Y)^{\tilde{c}}}$. Substituting equations \eqref{social_specific}-\eqref{social_specific4} in \eqref{c_eqn_social}, one can derive the following expressions,
\begin{align}
\label{c_eqn_social1}
\text{ln}\Big(\frac{\pi(X|\text{'}x\text{'})}{1-\pi(X|\text{'}x\text{'})}\Big)=   0.888 \times \text{ln}\Big(\frac{Pr(D_n=X|x) \ p(x|X)+Pr(D_n=X|y) \ p(y|X)}{Pr(D_n=X|x) \ p(x|Y)+Pr(D_n=X|y) \ p(y|Y)}\Big)
\end{align}
\begin{align}
\label{c_eqn_social2}
\text{ln}\Big(\frac{\pi(X|\text{'}y\text{'})}{1-\pi(X|\text{'}y\text{'})}\Big)=   0.888 \times \text{ln}\Big(\frac{[1-Pr(D_n=X|x)] \ p(x|X)+[1-Pr(D_n=X|y)] \ p(y|X)}{[1-Pr(D_n=X|x)] \ p(x|Y)+[1-Pr(D_n=X|y)] \ p(y|Y)}\Big)
\end{align}
where $\pi(X|.)$ is the subject's posterior after observing the neighbor's guess. To estimate the subject's belief about the neighbor's behavior, I assume  $\tilde{c}=0.888$, i.e., the subject's belief about the neighbor's posterior bias is correct. Then, I estimate $\tilde{\beta}$ by using a non-linear least square method to fit the data to equations \eqref{c_eqn_social1}-\eqref{c_eqn_social2}. The non-linear least square estimate for $\tilde{\beta}$ is 0.038 ($s.d. = 0.009$), which is significantly lower than the corresponding parameter for the individual reasoning error estimated in the first step, $\hat{\beta}=0.472 \ (s.d. = 0.018)$. This result suggests that the uncertainty in the social interaction causes the subjects to assign a lower response precision to their neighbor than their own. That is, subjects think their neighbors make more errors than they actually do.

\end{document}